\documentstyle[12pt,aaspp4]{article}
 
\begin{document}
 
\title{Image-Subtraction Photometry of Variable Stars in the Field
of the Globular Cluster NGC~6934\footnote{Based on observations
obtained with the 1.2 m Telescope at the F. L. Whipple Observatory of
the Harvard-Smithsonian Center for Astrophysics}}

\author{J. Kaluzny\altaffilmark{2}, A. Olech\altaffilmark{3}
and K. Z. Stanek\altaffilmark{4,5}}

\altaffiltext{2}{Copernicus Astronomical Center, ul. Bartycka 18,
00-716 Warsaw, Poland (jka@camk.edu.pl)}
\altaffiltext{3}{Warsaw University Observatory, Al.~Ujazdowskie~4,
00-478~Warsaw, Poland\\ (olech@sirius.astrouw.edu.pl)}
\altaffiltext{4}{Hubble Fellow}
\altaffiltext{5}{Harvard-Smithsonian Center for Astrophysics, 60 Garden
St., MS20, Cambridge, MA 02138, USA (kstanek@cfa.harvard.edu)}

\begin{abstract} 

We present CCD $BVI$ photometry of 85 variable stars from the field of
the globular cluster NGC~6934.  The photometry was obtained with the
image subtraction package ISIS. 35 variables are new identifications:
24 RRab stars, 5 RRc stars, 2 eclipsing binaries of W UMa-type, one SX
Phe star, and 3 variables of other types.  Both detected contact
binaries are foreground stars.  The SX Phe variable belongs most
likely to the group of cluster blue stragglers.  Large number of newly
found RR~Lyr variables in this cluster, as well as in other clusters
recently observed by us, indicates that total RR~Lyr population
identified up to date in nearby galactic globular clusters is
significantly ($>30\%$) incomplete.

Fourier decomposition of the light curves of RR Lyr variables was used
to estimate the basic properties of these stars.  From the analysis of
RRc variables we obtain a mean mass of $M=0.63\;M_\odot$, luminosity
$\log L/L_\odot=1.72$, effective temperature $T_{\rm eff}=7300$ and
helium abundance $Y=0.27$.  The mean values of the absolute magnitude,
metallicity (on Zinn's scale) and effective temperature for RRab
variables are: $M_V=0.81$, ${\rm [Fe/H]}=-1.53$ and $T_{\rm
eff}=6450$, respectively.  From the $B-V$ color at minimum light of
the RRab variables we obtained the color excess to NGC 6934 equal to
$E(B-V)=0.09\pm 0.01$.  Different calibrations of absolute magnitudes
of RRab and RRc available in literature were used to estimate apparent
distance modulus of the cluster: $(m-M)_{\rm V}=16.09\pm 0.06$. We
note a likely error in the zero point of the {\em HST}-based $V$-band
photometry of NGC~6934 recently presented by Piotto et al. Among
analyzed sample of RR Lyr stars we have detected a short period and
low amplitude variable which possibly belongs to the group of second
overtone pulsators (RRe subtype variables).

The $BVI$ photometry of all variables is available electronically via
{\tt anonymous ftp}.  The complete set of the CCD frames is available
upon request.

\end{abstract}


\section{Introduction}

NGC 6934 (${\rm RA}=20^h34^m$, ${\rm Decl.}=+7^\circ24'$, J2000.0) is
an intermediate-metallicity globular cluster.  In his catalog Harris
(1996) adopted for it ${\rm [Fe/H]}=-1.54 $.  The color magnitude
diagrams of this cluster were obtained by Harris and Racine (1973),
Brocato et al. (1998) and Piotto et al. (1999).  The cluster was
extensively searched for variable stars by Sawyer-Hogg \& Wehlau
(1980), who listed 51 variable stars, 50 of which were RR Lyr
variables. They have found mean periods for 45 RRab and 5 RRc stars
equal to $0.552\;$days and $0.294\;$days, respectively. These values
place NGC~6934 among Oosterhoff type I clusters.

\section{Observations and Reductions}

Observations analyzed in this paper were obtained as result of a
side-survey conducted during project DIRECT (Kaluzny et al. 1998a;
Stanek et al.~1998). The cluster was monitored with the 1.2~m
telescope at the F.L. Whipple Observatory (FLWO), where we used
"AndyCam" camera (Szentgyorgyi et al. 2000) containing Loral 2048$^2$
back-side illuminated CCD.  The pixel scale was $0.32\;arcsec/pixel$,
giving field of view roughly $11\times 11\;arcmin^2$. The monitored
field covers most of the cluster area as the tidal radius of NGC~6934
is estimated at $r=8.37\;arcmin$ (Harris 1996). The data were
collected from 1997 July 15 to September 23. The cluster was observed
early in the night when main targets of the DIRECT project, M31 and
M33, were located too far east to observe. Photometry of variable
stars presented in this paper is based on 78 $V$-band images, 22
$B$-band images and 21 $I$-band images, collected during 15
nights.\footnote{The complete list of exposures for the cluster and
related data files are available through {\tt anonymous ftp} on {\tt
cfa-ftp.harvard.edu}, in {\tt pub/kstanek/NGC6934} directory. Please
retrieve the {\tt README} file for instructions.}  For almost all
$V$-band images an exposure time was set to $450\;$sec and the median
value of seeing for that filter was $FWHM=1.7\;arcsec$.  Preliminary
processing of the CCD frames was done with the standard routines in
the IRAF CCDPROC package\footnote{IRAF is distributed by the National
Optical Astronomy Observatories, which is operated by the Association
of Universities for Research in Astronomy, Inc., under agreement with
the National Science Foundation}.

Initially we reduced our data by extracting profile photometry with
the DAOPHOT/ALLSTAR package (Stetson 1987, 1991). We followed a
procedure adopted by the DIRECT team, which is described in detail in
Kaluzny et al. (1998a). Inspection of derived data bases lead to
recovery of 50 out of 51 variables listed in Sawyer-Hogg \& Wehlau
(1980). The only unrecovered variable, star V15, was missed due to
fact that its images were overexposed on most of frames.  In addition
we identified 27 new variables located in the cluster field.

We attempted to improve quality of derived light curves by employing
image subtraction package ISIS.V2.1 (Alard \& Lupton 1998; Alard
2000)\footnote{ ISIS2 package can be down-loaded from
http://www.iap.fr/users/alard/package.html}.  It resulted not only in
better quality of photometry for already identified variables but also
allowed us to find six additional variable objects. We followed
prescription given in the ISIS.V2 manual to obtain differential light
curves expressed in ADU units. An additional step is needed to convert
ISIS light curves into magnitudes. This was accomplished by using
DAOPHOT/ALLSTAR based profile photometry derived from individual
images selected as templates (one image for every filter).  For every
variable its total flux registered on a template image was derived
based on its individual magnitude and an appropriate aperture
correction for a given frame. For some variables their profile
photometry turned out to be unreliable due to problems caused by
crowding in the innermost part of the cluster. For these stars we
decided to give up on transforming their ISIS based light curves into
magnitude units.  Specifically we transformed to magnitude units only
stars for which ALLSTAR returned profile photometry with $\sigma\leq
0.05$ and $CHI1\leq 3.0$.  In the following discussion we are using
solely light curves based on ISIS results.

Transformation from the instrumental magnitudes to the standard
$BVI_{c}$ system was accomplished based on observations of a few dozen
stars from several Landolt (1992) fields. These date were collected on
2 photometric nights.  The following relations were adopted for the
night of Sep 22/23, 1997:
\begin{eqnarray}
v=V+0.038\times (B-V) +0.116\times X +const\\
b=B-0.035\times(B-V)+0.198\times X +const\\
i=I-0.054\times (V-I) +0.052\times X +const
\end{eqnarray}
  
We estimate that total uncertainties (including uncertainty of
aperture corrections) of the zero points of cluster photometry should
not exceed $0.035\;$mag for all three filters used.

\section{Results}

In our search for variables in NGC 6934 we have identified 85 stars.
50 of them were previously known (Sawyer Hogg \& Wehlau 1980) and 35
are new discoveries. All previously known variables are RR Lyr stars
with five objects belonging to Bailey type c and 45 belonging to
Bailey type ab. Among newly identified variables we have detected 24
RRab stars, five RRc stars, two eclipsing W UMa-type stars, one SX Phe
star and three other objects.

Basic elements (coordinates, periods, peak-to-peak $V$ amplitudes,
intensity averaged $BVI$ magnitudes and types) for all variables from
our sample are presented in Table~1. We assigned names NV52-NV84 to
the newly identified objects. Transformation from rectangular
coordinates returned by DAOPHOT to the equatorial coordinates was
obtained based on positions of 40 stars from the USNO-A2 catalog
(Monet et al. 1996) identified in our field. Rectangular XY
coordinates listed in Table~1 are tied to the system used in the
Clement (1997) catalog.  For variables V1-V51 we adopted periods from
Clement (1997).  The exception from that rule are variables V1, V11,
V20, V23 and V45 for which we obtained revised periods based on our
data alone.  These new periods produce less scattered light curves
than old ones.  Periods for the newly identified variables NV52-NV86,
as well as for five other objects listed above, were derived using ORT
algorithm developed by Schwarzenberg-Czerny (1997).

\begin{figure}[p]
\plotfiddle{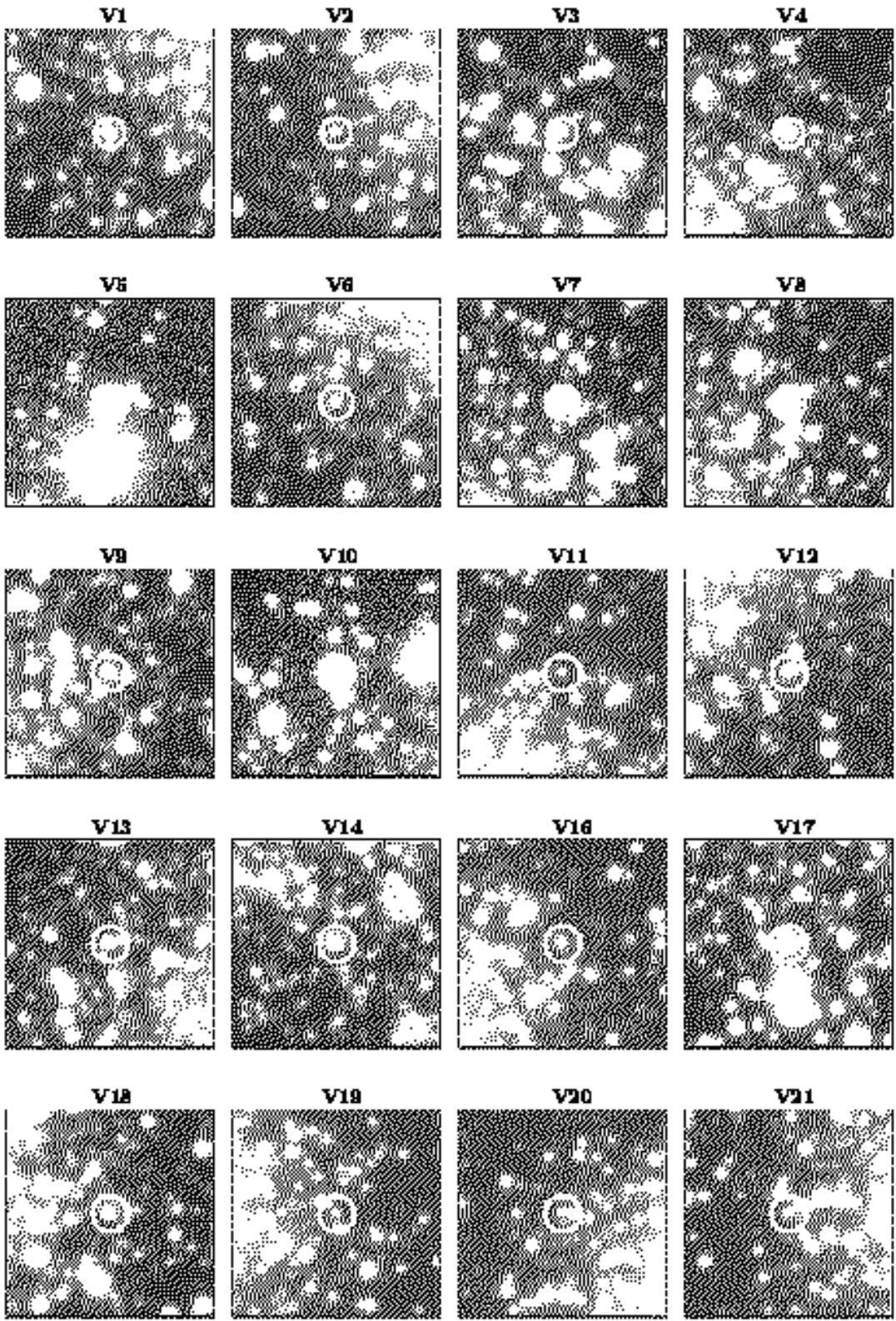}{19.5cm}{0}{93}{93}{-268}{-125}
\caption{Finder charts for NGC~6934 variables. Each chart is
$40$~arcsec wide with North up and East to the left.}
\label{fig:fcharts}
\end{figure}
 
\addtocounter{figure}{-1}
\begin{figure}[p]
\plotfiddle{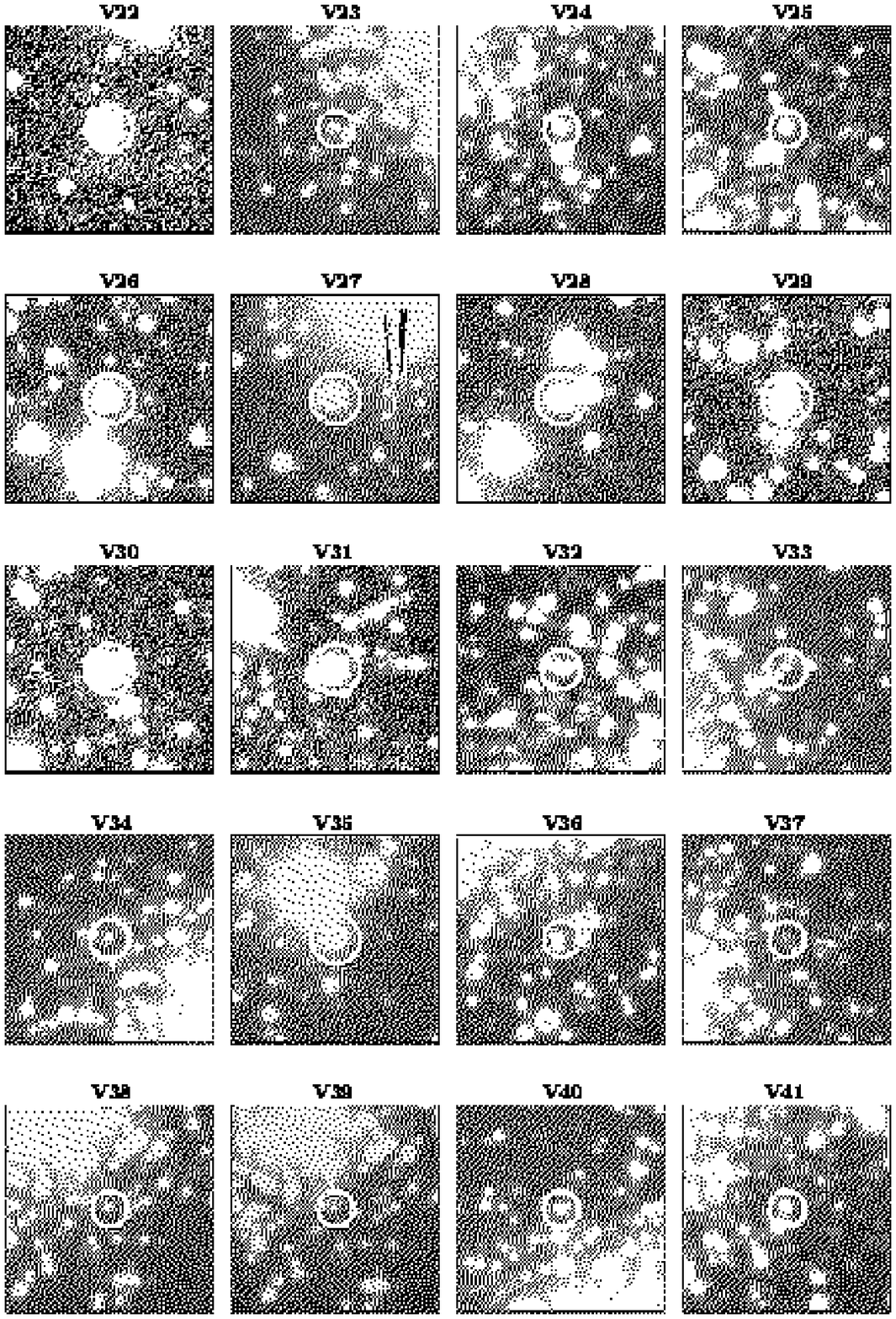}{19.5cm}{0}{93}{93}{-268}{-125}
\caption{Continued.}
\end{figure}
 
\addtocounter{figure}{-1}
\begin{figure}[p]
\plotfiddle{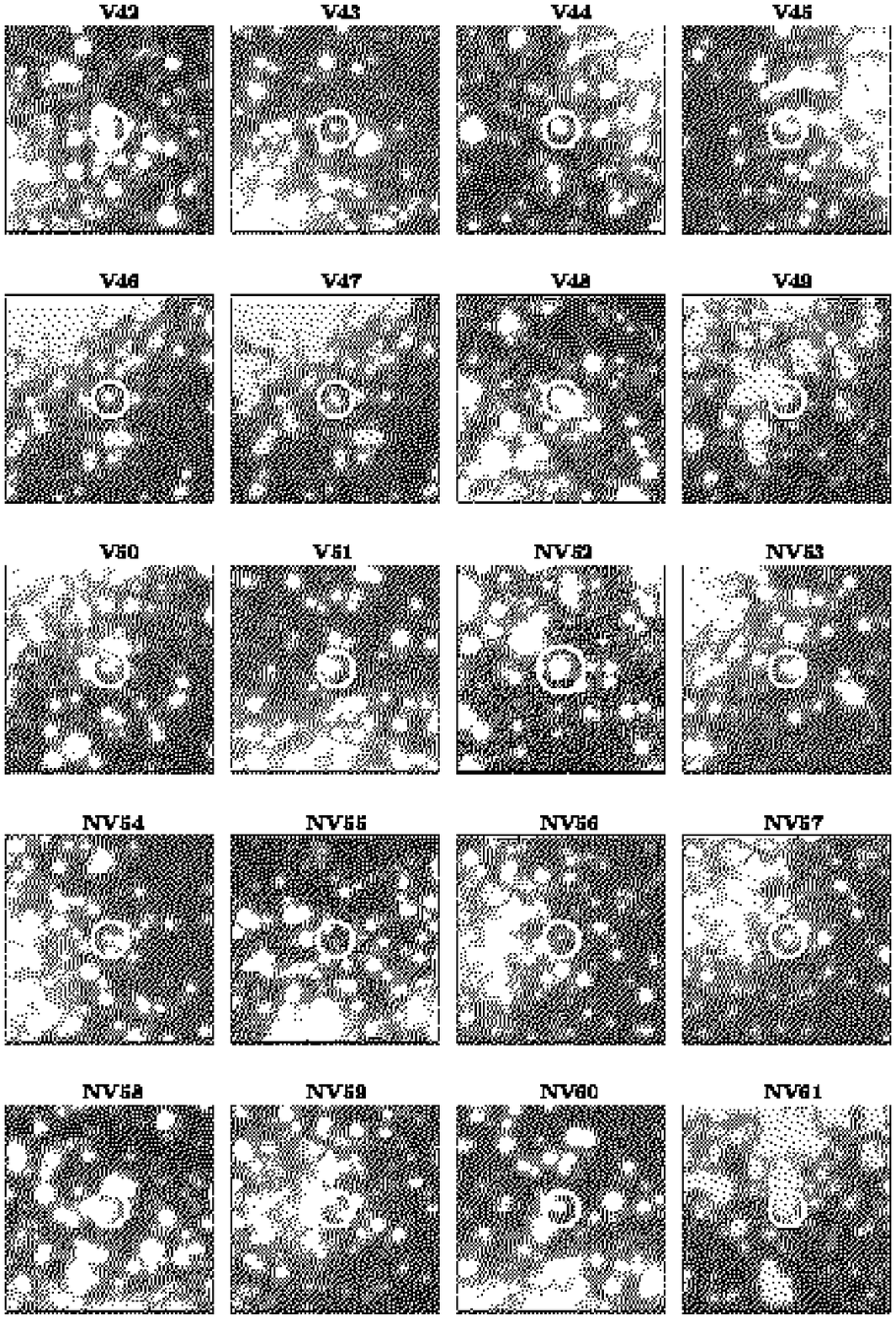}{19.5cm}{0}{93}{93}{-268}{-125}
\caption{Continued.}
\end{figure}

\addtocounter{figure}{-1}
\begin{figure}[p]
\plotfiddle{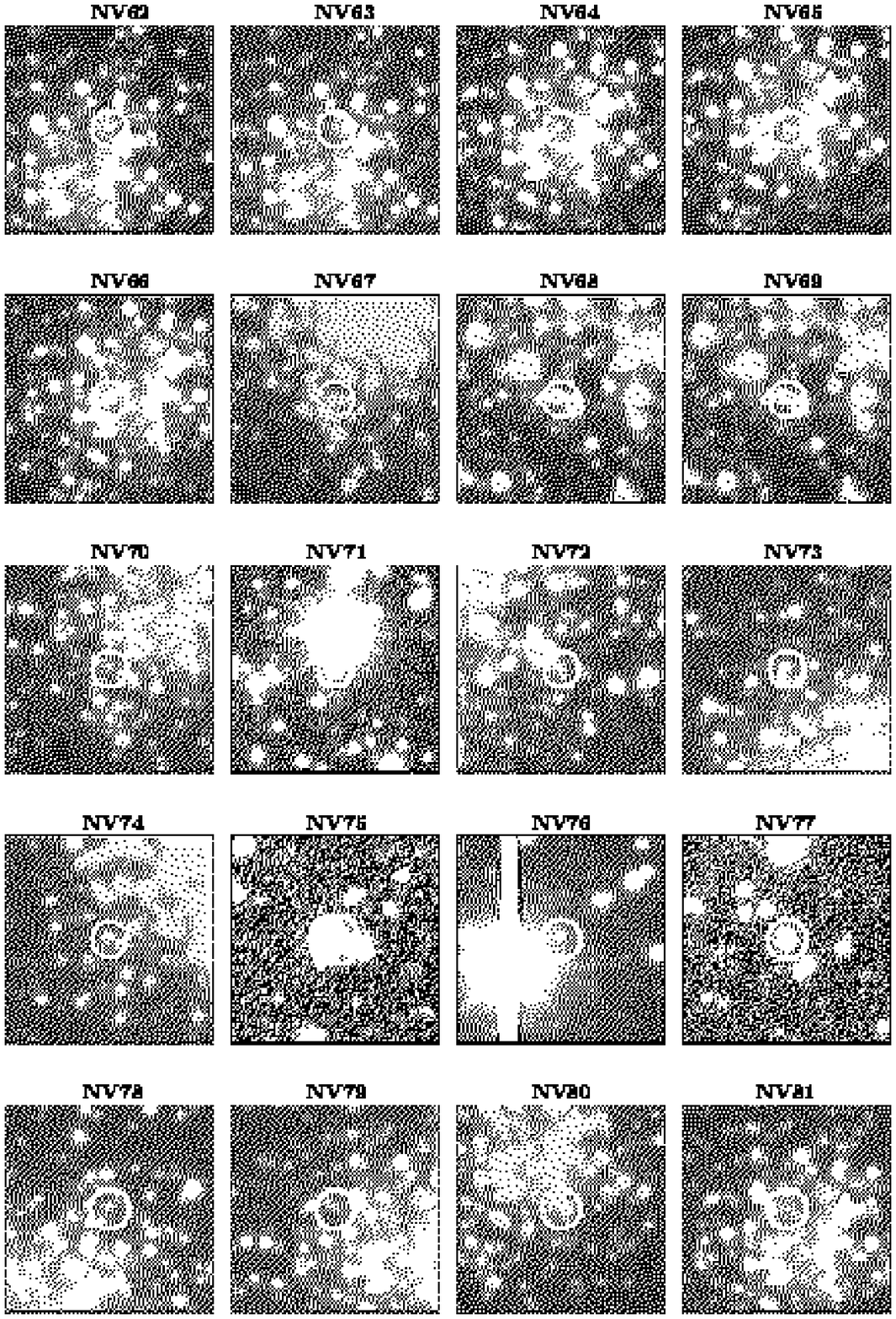}{19.5cm}{0}{93}{93}{-268}{-125}
\caption{Continued.}
\end{figure}

\addtocounter{figure}{-1}
\begin{figure}[t]
\plotfiddle{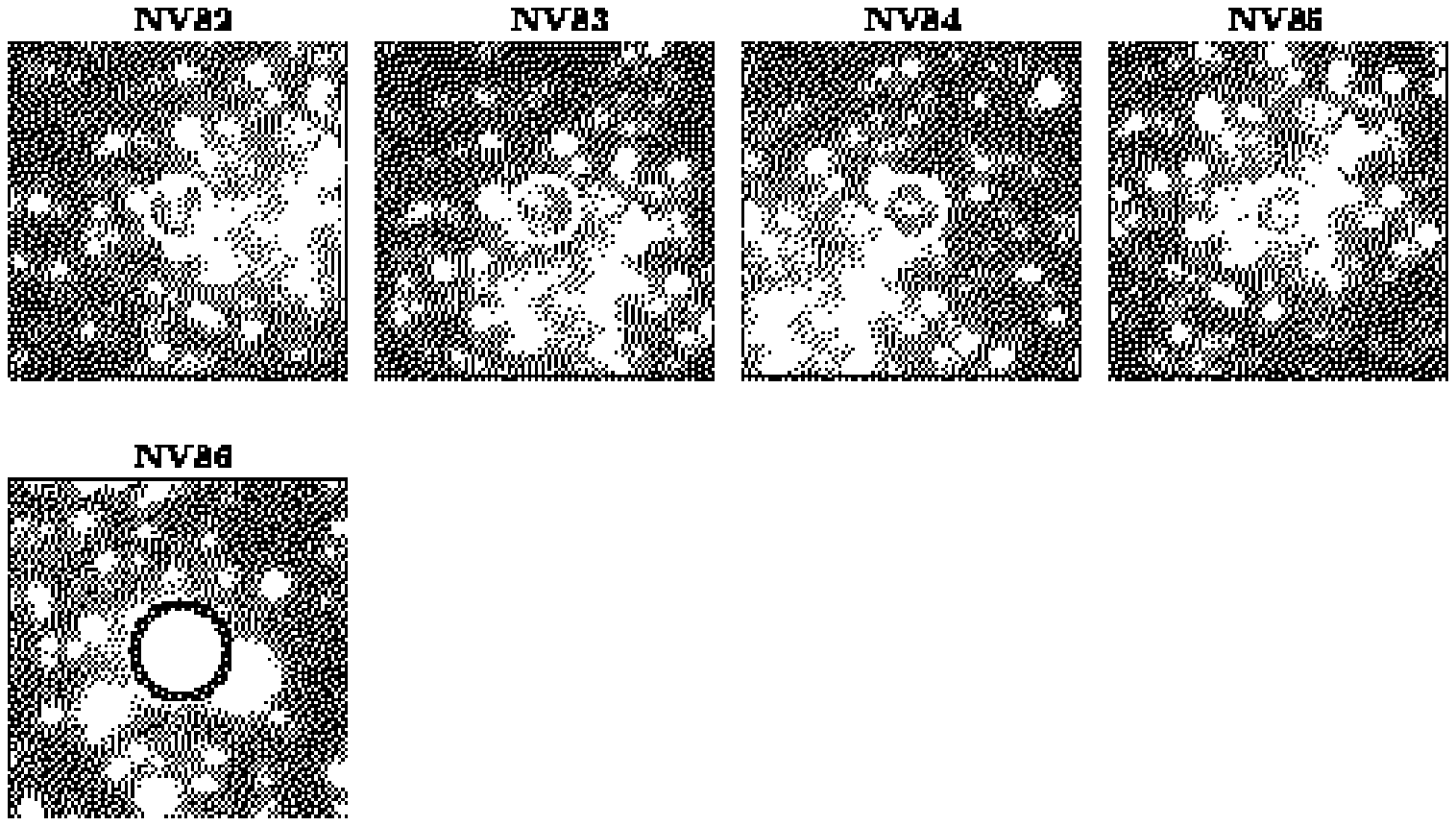}{6.5cm}{0}{93}{93}{-268}{-495}
\caption{Continued.}
\end{figure}

\begin{figure}[p]
\plotfiddle{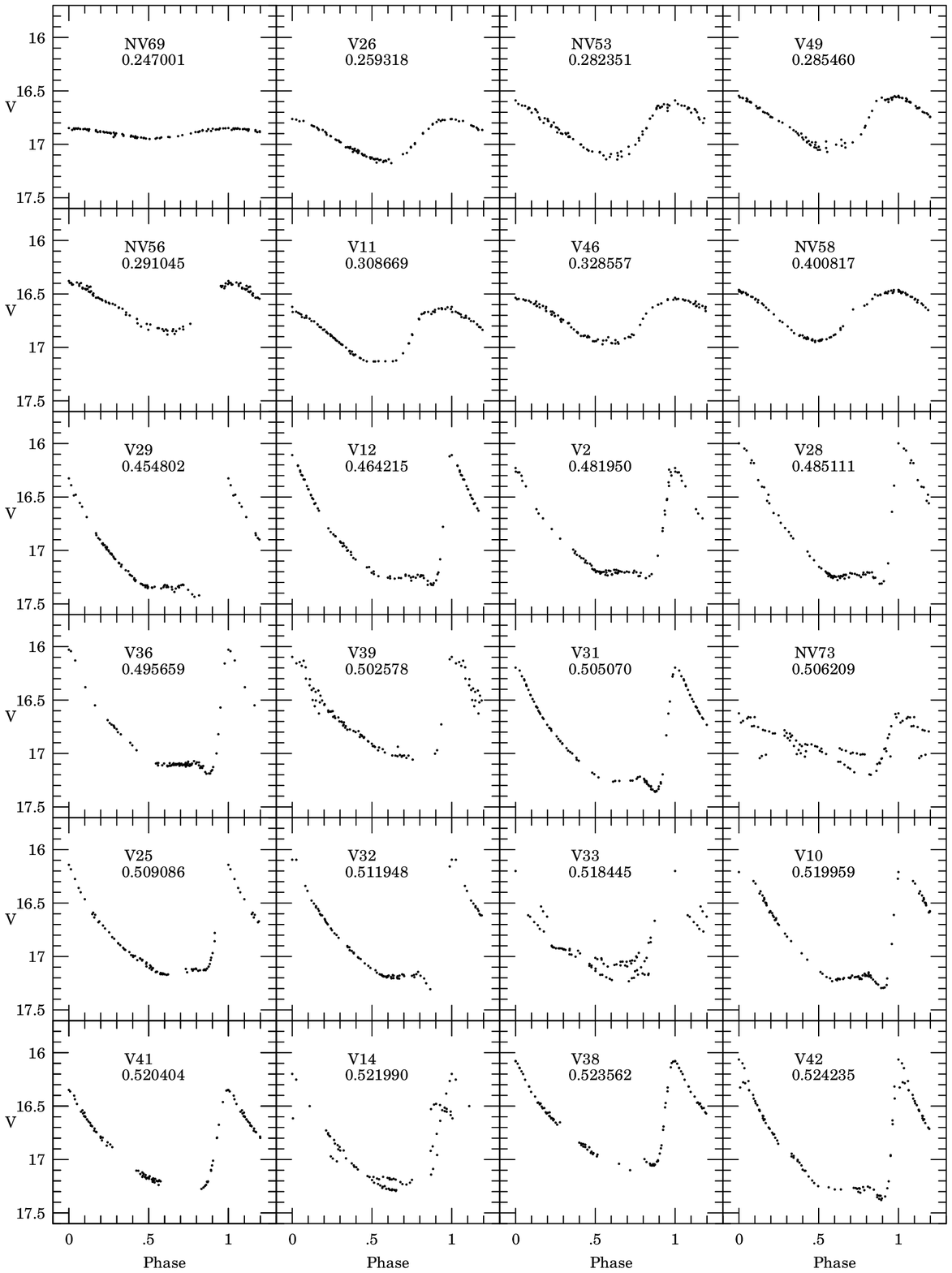}{19.5cm}{0}{93}{93}{-288}{-100}
\caption{$V$-band light curves of variables for which ISIS photometry
was transformed to magnitudes.  The stars are plotted according to the
increasing period.}
\label{fig:lcurs}
\end{figure}

\addtocounter{figure}{-1}
\begin{figure}[p]
\plotfiddle{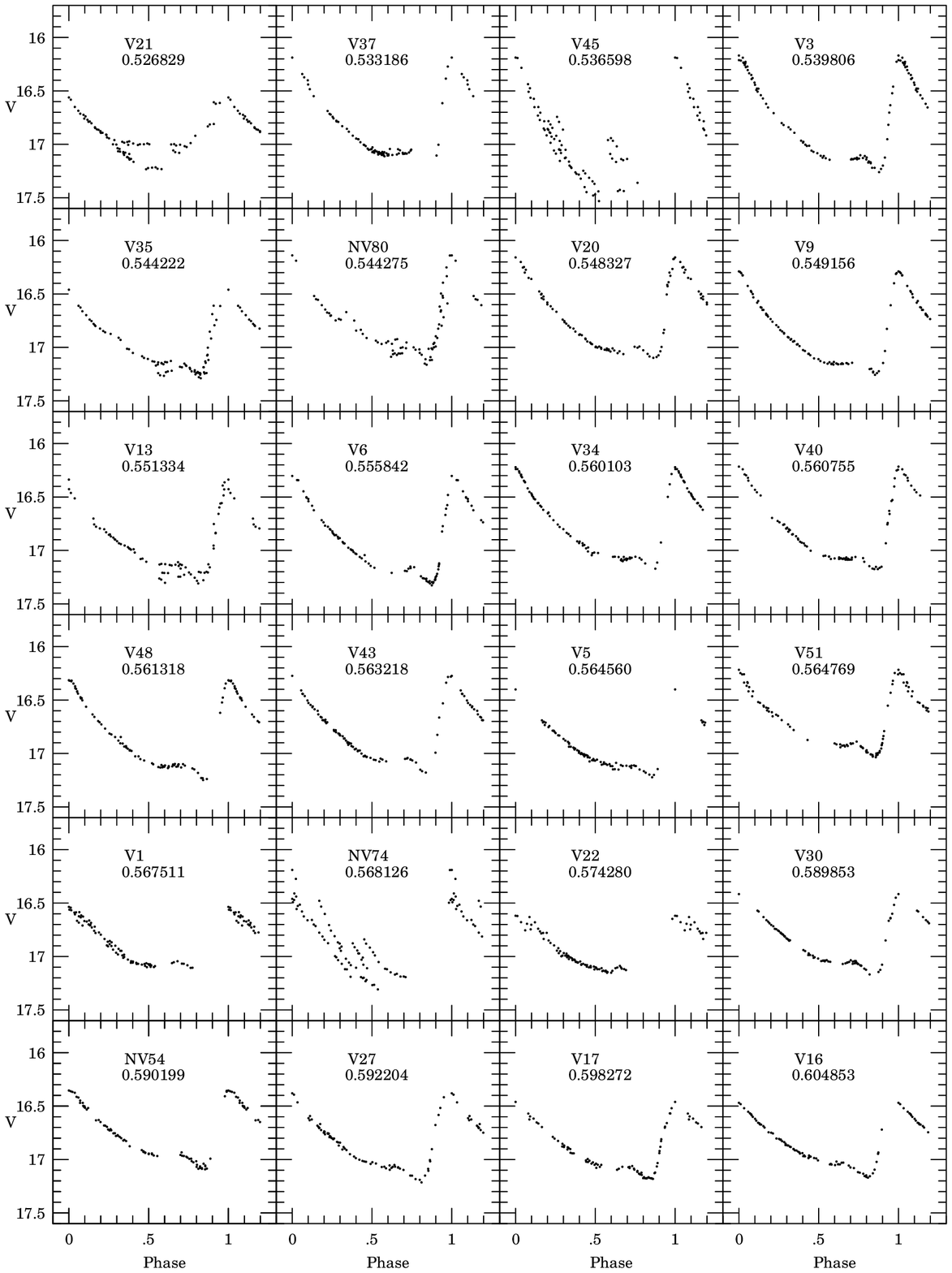}{19.5cm}{0}{93}{93}{-288}{-100}
\caption{Continued.}
\end{figure}

\addtocounter{figure}{-1}
\begin{figure}[p]
\plotfiddle{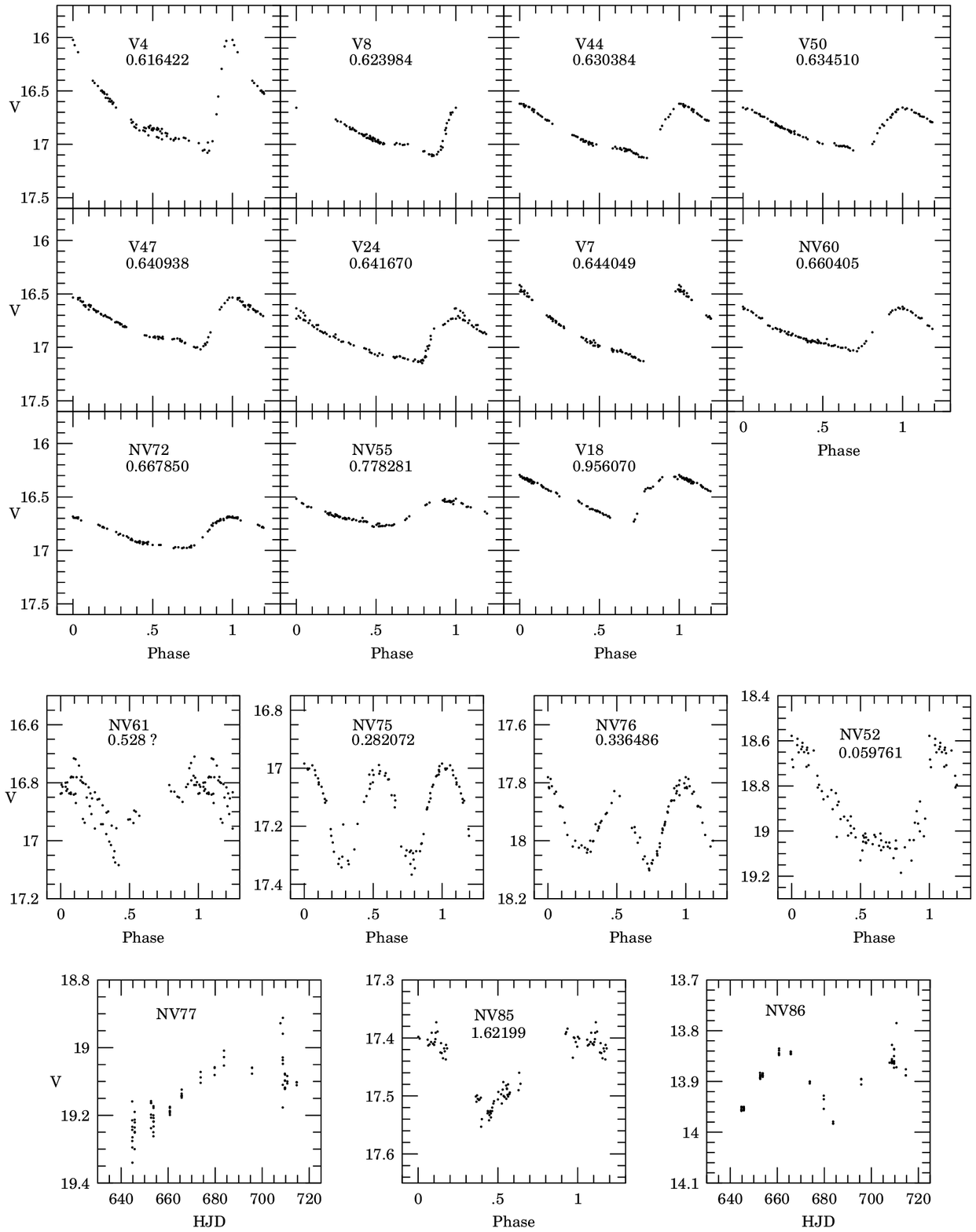}{19.5cm}{0}{93}{93}{-288}{-100}
\caption{Continued.}
\end{figure}

\begin{figure}[p]
\plotfiddle{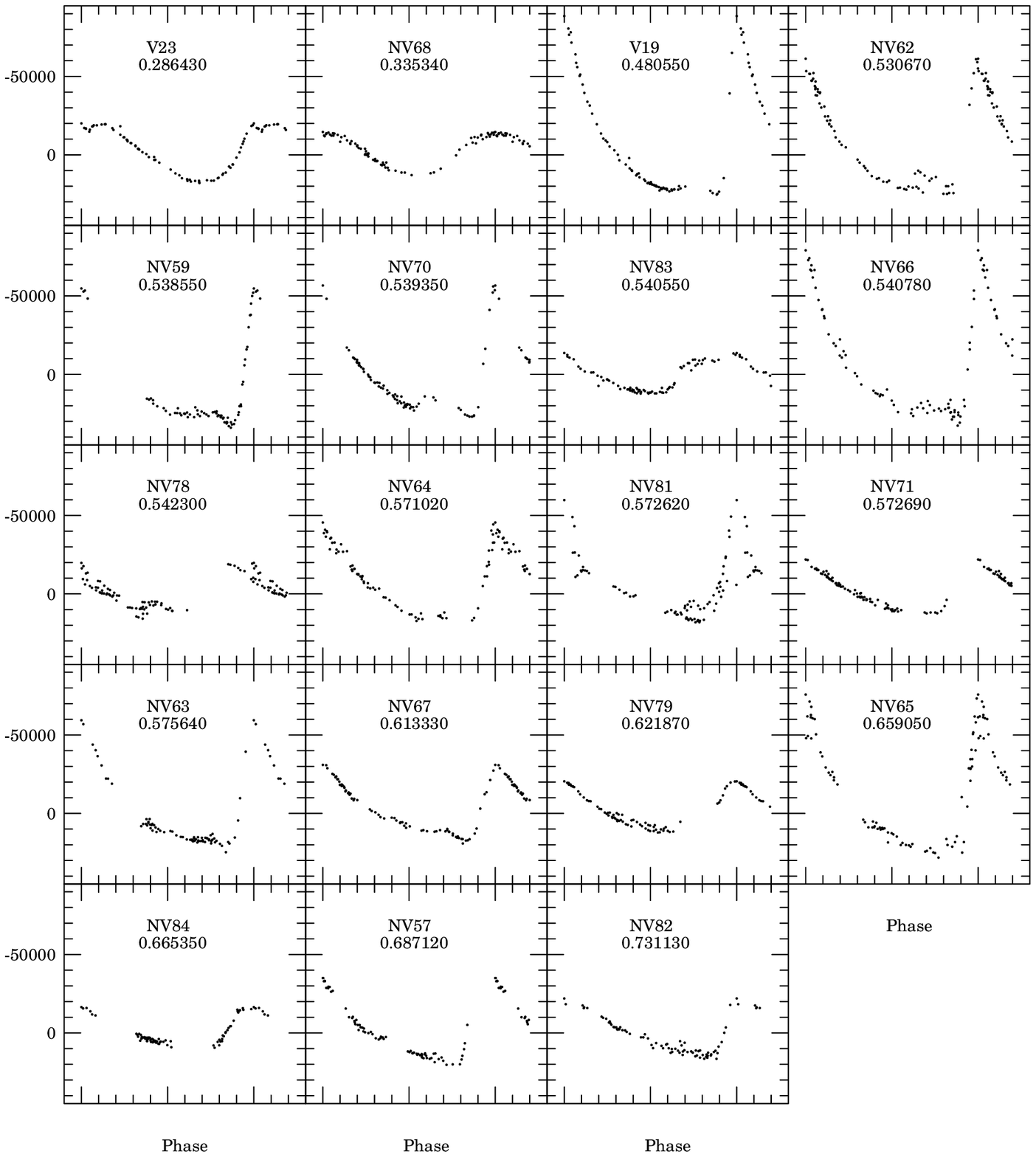}{16.5cm}{0}{100}{100}{-295}{-210}
\caption{$V$-band light curves for variables whose ISIS photometry was
retained in differential counts units.  The stars are plotted
according to the increasing period.}
\label{fig:lalard}
\end{figure}

Finding charts for all variables discussed in this paper are shown in
Fig.~1. Each chart is $40$~arcsec wide with North up and East to the
left.  In Fig.~2 we present $V$-band light curves for variables whose
ISIS based photometry was transformed into magnitude units.  Light
curves which were left in the differential counts units are shown in
Fig.~3.

In Fig.~4 we show the map of the inner part of NGC~6934 with positions
of the variables marked by open circles. The dotted line marks the
central part of the cluster with $r=0.9$~arcmin. In this circle, for
clarity, we do not plot the constant stars. Note that we find variable
stars all the way to the center of the cluster.

In Fig.~5 we present a $V/B-V$ and $V/V-I$ color-magnitude diagrams
derived for the field of NGC~6934. For every filter presented
photometry was obtained by averaging ALLSTAR results obtained from
5-11 frames.  Identified variable stars are marked with special
symbols in Fig.~5.

We note that our survey increased by almost 60\% population of known
RR~Lyr variables in NGC~6934. Similar results were recently reported
for other well studied clusters like M5 (Olech et al. 1999), M55
(Olech at al.  1999a) or NGC~6362 (Mazur, Kaluzny \& Krzeminski 1999).
It seems that contrary to some claims (eg. Suntzeff, Kinman \& Kraft
1991), the sample of RR~Lyr variables identified up to date in
galactic globular clusters is significantly incomplete.

\subsection{RR Lyr Variables}

The majority of variable stars identified in NGC 6934 are of RR Lyr
type.  Among them there are 10 RRc stars and 69 RRab stars. Such a
large ratio of RRab to RRc variables is common for Oosterhoff type I
globular clusters. According to Smith (1995) the average ratio between the
number of RRc stars and all RR Lyr variables $n(c)/n(c+ab)$ in
Oosterhoff type I clusters is equal to 0.17 and for Oosterhoff type II
clusters it is equal to  0.44.  For NGC 6934 we obtained $n(c)/n(c+ab)=0.13$.

\begin{figure}[p]
\plotfiddle{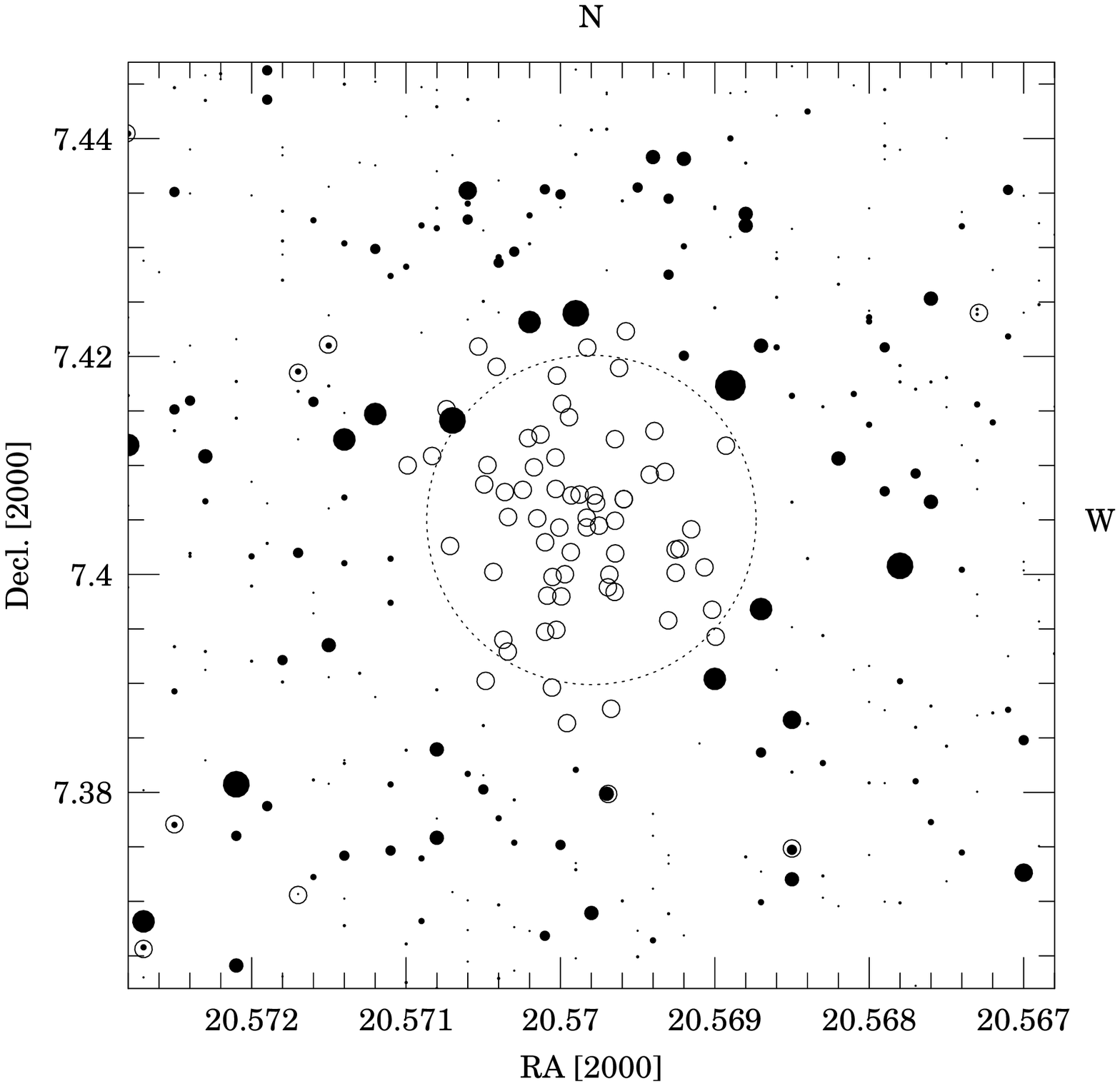}{8.7cm}{0}{55}{55}{-178}{-95}
\caption{The map of the inner NGC~6934 with positions of the variables
marked by open circles. The dotted line marks central part of the 
cluster with $r=0.9$~arcmin. In that part, for clarity, we do not 
plot the constant stars.}
\label{fig:chart}

\plotfiddle{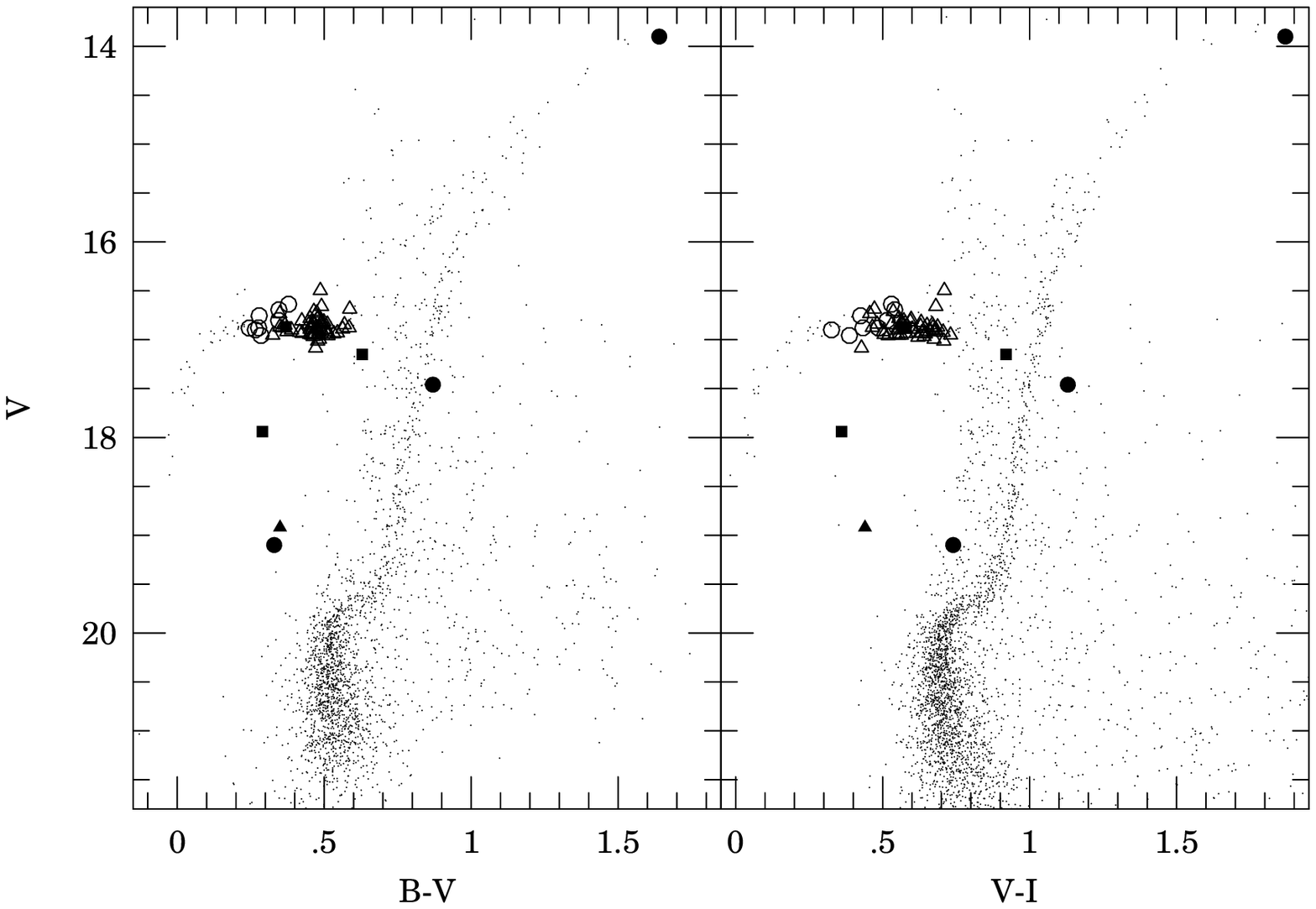}{8.7cm}{0}{80}{80}{-260}{-305}
\caption{The $(V,B-V)$ and $(V,V-I)$ color-magnitude diagrams for NGC
6934. Open circles and triangles denote the RRc and RRab stars,
respectively. Filled squares denote W~UMa stars, filled triangle
denotes SX~Phe variable and filled circles denote other variables.}
\label{fig:cmd}
\end{figure}

The periods of RRab variables from NGC 6934 are between 0.4548 and
$0.956\;$days with the mean value of $0.574\;$days. Periods of RRc
stars are between $0.247$ and $0.4008\;$days with the mean at $0.310
\;$days. According to Smith (1995) the mean periods of RRc and RRab
stars are $0.32\;$days and $0.55\;$days in Oosterhoff type I clusters
and $0.37\;$days and $0.64\;$days in Oosterhoff type II clusters.

\begin{figure}[t]
\plotfiddle{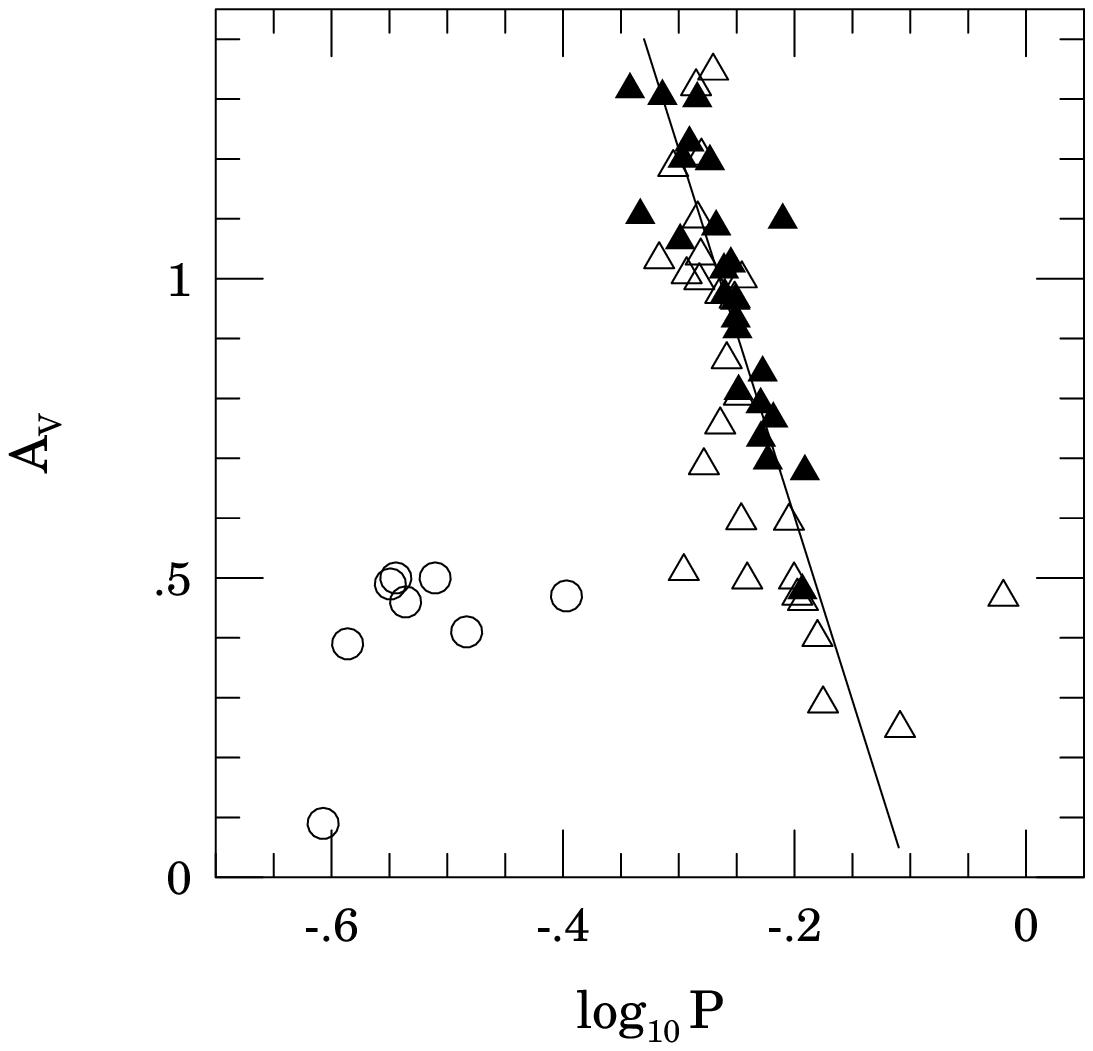}{8.7cm}{0}{97}{97}{-270}{-255}
\caption{Period-amplitude diagram for RRab stars with $D_m<3$ (solid
triangles), RRab stars wit $D_m>3$ (open triangles) and RRc stars
(open circles). The solid line represents a linear fit to RRab
variables in M3 (Kaluzny et al.~1998).}
\label{fig:pampl}
\end{figure}

The values of the peak-to-peak amplitudes $A_V$ and periods $P$
presented in Table~1 are used to plot the
period-amplitude ($\log P-A_{V}$) diagram shown in Fig.~6. Open circles
denote RRc stars, filled triangles RRab stars with $D_m<3$ (for
definition of $D_m$ see Kovacs \& Kanbur 1998) and open triangles RRab
variables with $D_m>3$. The solid line represents a linear fit to RRab
variables in M3 (Kaluzny et al. 1998) an another Oosterhoff type I
globular cluster. One can see that RRab variables from NGC 6934 follow
closely the linear relation for M3. It is not surprising because both
clusters have very similar metallicities.

We fitted our $V$-band light curves with Fourier sine series of the
form:
\begin{equation}
V=A_0+\sum^{6}_{j=1}A_j\cdot\sin(j\omega t + \phi_j),
\end{equation}
where $\omega=2\pi/P$. To find the values of $\omega$, $A_j$ and
$\phi_j$ we employed the method developed by Schwarzenberg-Czerny
(1997) and Schwarzenberg-Czerny \& Kaluzny (1998).  The elements of
the Fourier decomposition of the light curves are used in the
following subsections for determination of the physical parameters for
RR Lyr variables from NGC 6934.

\subsubsection{A Fourier Analysis of the RRc Stars}

It was demonstrated by Simon \& Clement (1993) that Fourier
decomposition of light curves of RRc variables is a very useful
technique for determining physical parameters of these stars. The
equations of Simon \& Clement (1993) are:
\begin{equation} 
\log M = 0.52\log P_1 - 0.11\phi^*_{31} + 0.39 
\end{equation}
\begin{equation}
\log L = 1.04\log P_1 - 0.058\phi^*_{31} + 2.41
\end{equation}
\begin{equation}
\log T_{\rm eff} = 3.265 - 0.3026\log P_1 - 0.1777\log M + 0.2402\log L
\end{equation}
\begin{equation}
\log Y = -20.26 + 4.935\log T_{\rm eff} - 0.2638\log M+ 0.3318\log L
\end{equation}
where $M$ is the mass of the star in solar units, $P_1$ is the first
overtone pulsation period in days, $L$ is the luminosity in solar
units, $T_{\rm eff}$ is the effective temperature, $Y$ is the relative
helium abundance and $\phi^*_{31}=\phi^*_3-3\phi^*_1$. The phases
marked by asterisk are obtained from a cosine Fourier series (used by
Simon \& Clement 1993) and differ from our phases which were obtained
from a sine series (cf. Equation 4). For $\phi_{31}$ we have
$\phi_{31}=\phi^*_{31}+\pi$.

From Eqs. (5)--(8) we computed masses, luminosities, effective
temperatures, relative helium abundances and absolute magnitudes of
RRc stars in NGC 6934. These are presented in Table 2 together with
the values of $A_0$, $A_1$, $\phi_{21}$ and $\phi_{31}$.  The errors
presented in Table 2 are calculated from the formal errors of the
Fourier coefficients using the error propagation law.
 
Using the formula of Kov\'acs (1998) we can also compute the absolute
magnitude $M_V^{Ko}$:
\begin{equation}
M_V^{Ko}=1.261-0.961 P_1-0.004\phi_{21}-4.447 A_4.
\end{equation}
The above equation is calibrated using luminosities derived with the
Baade-Wesselink method. It implies relatively faint absolute
magnitudes of RR Lyr stars. We decided also to use the values of $\log
L/L_\odot$ of RRc stars for computing the other way absolute visual
magnitudes of RRc stars in NGC 6934. The values of $M_V$ (presented in
the last column of Table 2) were calculated assuming a value of 4.70
for $M_{\rm bol}$ of the Sun and using the bolometric correction ${\rm
BC}=0.06{\rm [Fe/H]}+0.06$ adopted after Sandage \& Cacciari (1990).

From a sample of 10 RRc stars detected in NGC 6934 we excluded two
variables for which only ISIS photometry is available. Four out of the
remaining eight stars have errors of $\phi_{31}$ larger than 0.2 and
these objects were also excluded from further analysis. For the four
retained objects the mean values of the mass, luminosity, effective
temperature and helium abundance are $0.63\pm 0.06~M_\odot$, $\log
L/L_\odot=1.72\pm 0.01$, $T_{\rm eff}=7290\pm 27$, and $Y=0.27\pm
0.01$, respectively.

\subsubsection{Variable NV69 -- a second overtone pulsator?}

In recent years several authors suggested that RRc stars with periods
from the range $0.20-0.28\;$days may be in fact RRe variables, i.e. RR
Lyrae stars pulsating in the second overtone. Walker \& Nemec (1996)
found four such candidates in the globular cluster IC 4499. Other
authors (Alcock et al, 1996; Olech 1997) analyzing the period
distributions of RR Lyr variables in LMC and in the Galactic Bulge found
three peaks at periods 0.58, 0.34 and 0.28 days, corresponding to the
RRab, RRc and possibly to RRe stars, respectively.  Most recently Kiss
et al.~(1999) presented the detailed study of V2109 Cygni - an
ultra-short, small amplitude RRc star. They concluded that it occupies
different regions on the period-amplitude diagrams than other RRc
stars and thus is indeed RRe star.  On the other hand Kov\'acs (1998a)
presented some arguments against observational evidence for presence
of RRe stars with $<P>\approx 0.28\;$days.

The variable NV69 is the shortest period and lowest amplitude RR~Lyr
star in NGC~6934. On the period-amplitude relation shown in Fig.~6 it
occupies completely different location than other RRc stars. We also
constructed other diagrams with $R_{21}-\log P$, $\phi_{21}-\log P$
and $\phi_{31}-\log P$ relations, where $R_{21}=A_2/A_1$.  These
relations are shown in Fig.~7, where RRc variables are plotted with
open circles and RRab variables with solid triangles.  Solid circle
denotes the variable NV69 and solid square corresponds to V2109
Cygni. Due to the nearly sinusoidal shape of light curves of RRc stars
their phases $\phi_2$ and $\phi_3$ are often determined with low
precision what leads to large uncertainties of $\phi_{21}$ and
$\phi_{31}$ estimations. Thus NV69 and V2109 Cygni within error bars
may lay close to the whole group of RRc stars in $\phi_{21}-\log P$
and $\phi_{31}-\log P$ plots. Fortunately the errors of amplitudes
$A_1$ and $A_2$ are relatively small and therefore we can determine
$R_{21}$ with good precision. On the $R_{21}-\log P$ plot NV69 and
V2109 Cyg are located far away from RRc stars what supports hypothesis
that they are the second overtone pulsators and belong to RRe
variables.

A possible argument against classifying V69 as RRe variables comes
from theoretical models published by Bono et al. (1997).  These models
predict that for periods shorter than about 0.33 days, the amplitude
of pulsations observed for RRc stars should decrease with decreasing
period (the period-amplitude relation becomes parabolic). However,
decrease of period is accompanied by increase of $T_{eff}$ and
consequently RRc stars with very small amplitudes are expected to be
bluer than their counterparts with larger amplitudes. We may comment
that average color (B-V) observed for V69 is comparable to colors of 4
other RRc stars from NGC~6934. Specifically we obtained $<B-V>$ equal
to 0.25, 0.28, 0.27 and 0.28 for V11, V26, V46 and V53,
respectively. For V69 we measured $<B-V>=0.27$.

\subsubsection{RRab variables}

Kov\'acs \& Jurcsik (1996, 1997, and references quoted therein) have
extended the Fourier analysis of Simon \& Clement (1993) into RRab
stars. They derived the formulae that connect the periods, amplitudes
and phases of RRab stars with their physical parameters such as
absolute magnitude, metallicity, intrinsic colors and
temperatures. These equations are:
\begin{equation}
{\rm [Fe/H]} = -5.038 - 5.394 P_0 + 1.345\phi_{31}
\end{equation}
\begin{equation}
M_V = 1.221 - 1.396 P_0 - 0.477A_1 + 0.103\phi_{31}
\end{equation}
\begin{equation}
V_0-K_0=1.585 + 1.257 P_0 - 0.273A_1 - 0.234\phi_{31} + 0.062\phi_{41}
\end{equation}
\begin{equation}
\log T_{\rm eff} = 3.9291 - 0.1112(V_0-K_0) - 0.0032{\rm [Fe/H]}
\end{equation}
where $\phi_{41}=\phi_4-4\phi_1$ (cf. equation 4).

The above equations are valid only for RRab stars with regular light
curves, i.e. variables with a deviation parameter $D_m$ smaller than 3
(see Kov\'acs \& Kanbur 1998 for definition of $D_m$).  Table 3
summarizes the results obtained using Equations (10)-(13). This table
does not contain the variable NV61 due to its poor light curve and
uncertain period.  The parameters such as $\phi_{31}$, $\phi_{41}$,
$M_V$, $\Delta M_V$, [Fe/H] and $\Delta {\rm [Fe/H]}$ are listed only
for stars with $D_m<5$. The mean values of the absolute magnitude,
metallicity and effective temperature for 24 stars with $D_m<3$ are
$M_V=0.81\pm 0.01$, ${\rm [Fe/H]}=-1.31\pm 0.04$ and $T_{\rm
eff}=6455\pm 18$, respectively.

One can see that the value of [Fe/H] derived from Equation~10 differs
by about $0.2\;dex$ from the value adopted by Harris (1996). On the
other hand we should remember that the metallicity computed from
Equation (10) is in the scale of Jurcsik (1995) which is connected
with scale of Zinn \& West (1984) by the formula:
\begin{equation}
{\rm [Fe/H]_{Jurcsik}} = 1.431{\rm [Fe/H]_{ZW}}+0.880
\end{equation}
and thus ${\rm [Fe/H]}=-1.31$ on the Jurcsik's scale corresponds to
${\rm [Fe/H]}=-1.53$ on the Zinn \& West scale. Therefore our final
determination of metallicity of RR Lyr from NGC 6934 on Zinn and West
(1984) scale is ${\rm [Fe/H]}=-1.53\pm 0.03$. This result agrees very
well with the value adopted by Harris (1996; ${\rm[Fe/H]}=-1.54$) in
his catalog of globular clusters.

\subsection{Reddening of NGC 6934}

Previous determinations of reddening for NGC~6934 range from
$E(B-V)=0.05$ (Piotto et al. 1999) to $E(B-V)=0.20$ (Harris \& Racine
1973). The reddening map of Schlegel et al. (1999) gives $E(B-V)=0.11$
for the cluster position.  We can provide an independent estimate of
reddening for NGC~6934 using the method developed originally by
Preston (1964) and Sturch (1966). Following Blanco (1992) for RRab
variables the interstellar reddening can be estimated from the
formula:
\begin{equation}
E(B-V)=<B-V>_{\Phi(0.5-0.8)}+0.01222\Delta S 
- 0.00045(\Delta S)^2 - 0.185P -0.356
\end{equation}
where $<B-V>_{\Phi(0.5-0.8)}$ is the observed mean color in the
0.5-0.8 phase interval, $P$ is the fundamental period and $\Delta S$
is Preston's metallicity index. Based on the globular cluster
metallicity scale adopted by Zinn and West (1984) Suntzeff et
al. (1991) derived the following $\Delta S - {\rm [Fe/H]}$ relation:
\begin{equation}
{\rm [Fe/H]} = -0.408 - 0.158\Delta S
\end{equation}                                                 

From our sample of RRab variables we have selected 39 stars for which
with at least four determinations of $B-V$ color in the phase interval
0.5--0.8 are available.  For these stars we have used Equation (15)
and derived $E(B-V)=0.091\pm 0.006$ for the assumed $[{\rm
Fe/H}]=-1.54$.

\subsection{Distance Modulus}

Recent determination of distance modulus for NGC~6934 was published by
Piotto et al. (1999). They used an observed $V$ magnitude level of the
Zero Age Horizontal Branch ($V_{ZAHB}$) as a distance indicator and
obtained for the cluster $(m-M)_{V}=16.37$.  That estimate is based on
a measured value $V_{ZAHB}=17.05\pm 0.04$ and an adopted
$M_{V}^{ZAHB}=0.68$. The value of $V_{ZAHB}$ obtained for NGC~6934 by
Piotto et al. (1999) is noticeably fainter than earlier estimates
available in literature\footnote{Piotto et al. (1999) assume that peak
of the magnitude distribution of HB stars is equivalent to the level
of the ZAHB. They argue that such approach is justified by a fact that
globular clusters HB stars spent most of their evolution on the
ZAHB.}. Harris \& Racine (1973) used their photographic data to obtain
$V_{HB}=16.82\pm 0.02$ while Brocato et al. (1996) listed
$V_{HB}=16.90\pm 0.05$ based on the CCD data. To clarify this apparent
discrepancy we used our photometry to derive yet another estimate of
the horizontal branch level for the cluster. A histogram showing
distribution of $V$ magnitudes for non-variable HB stars from our
sample with $0.1\leq (B-V)\leq 0.7$ (the same color interval was used
by Piotto et al. (1999)) is presented in Fig.~8. The peak of that
histogram is located at $V_{HB}=16.86\pm 0.05$.

\begin{figure}[p]
\caption{Amplitude ratio and Fourier phase differences as
a function of period. RRc variables are plotted with open circles and
RRab variables with solid triangles. Solid circle denotes the variable
NV69 and solid square denotes V2109 Cygni.}
\plotfiddle{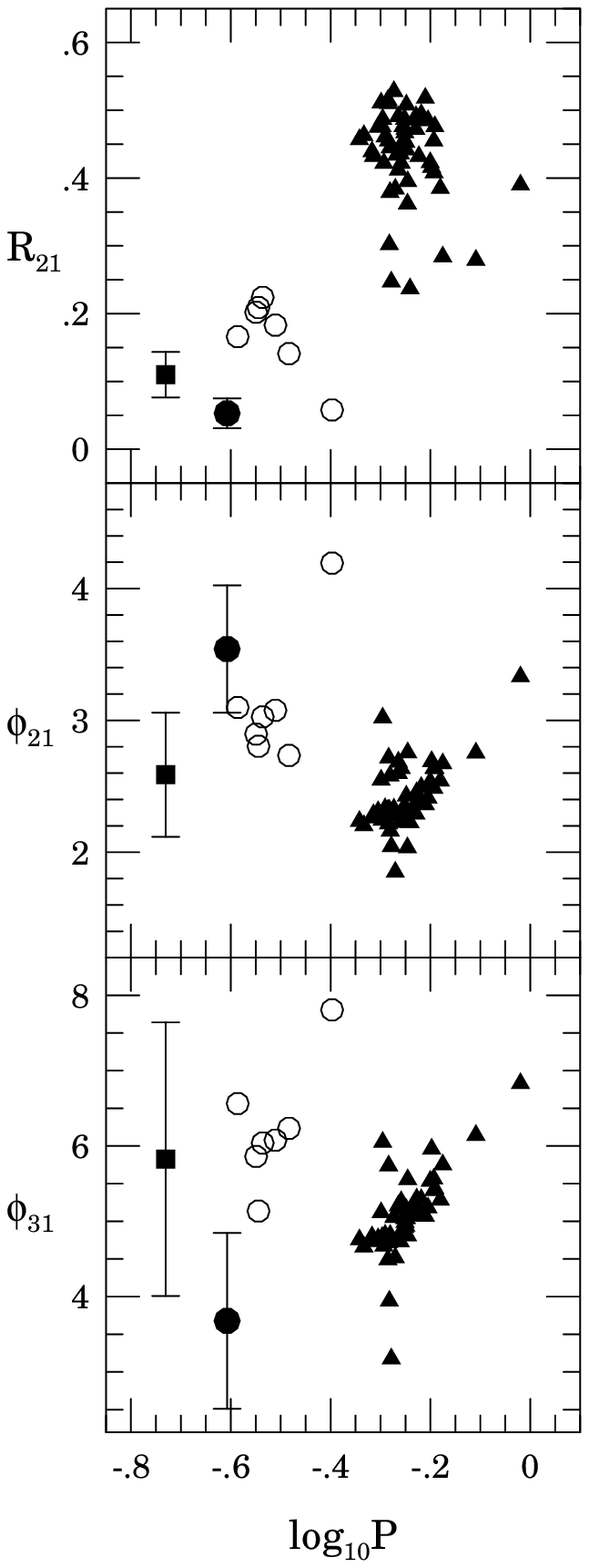}{8.7cm}{0}{90}{90}{-380}{-385}
\label{fig:fourier}

\plotfiddle{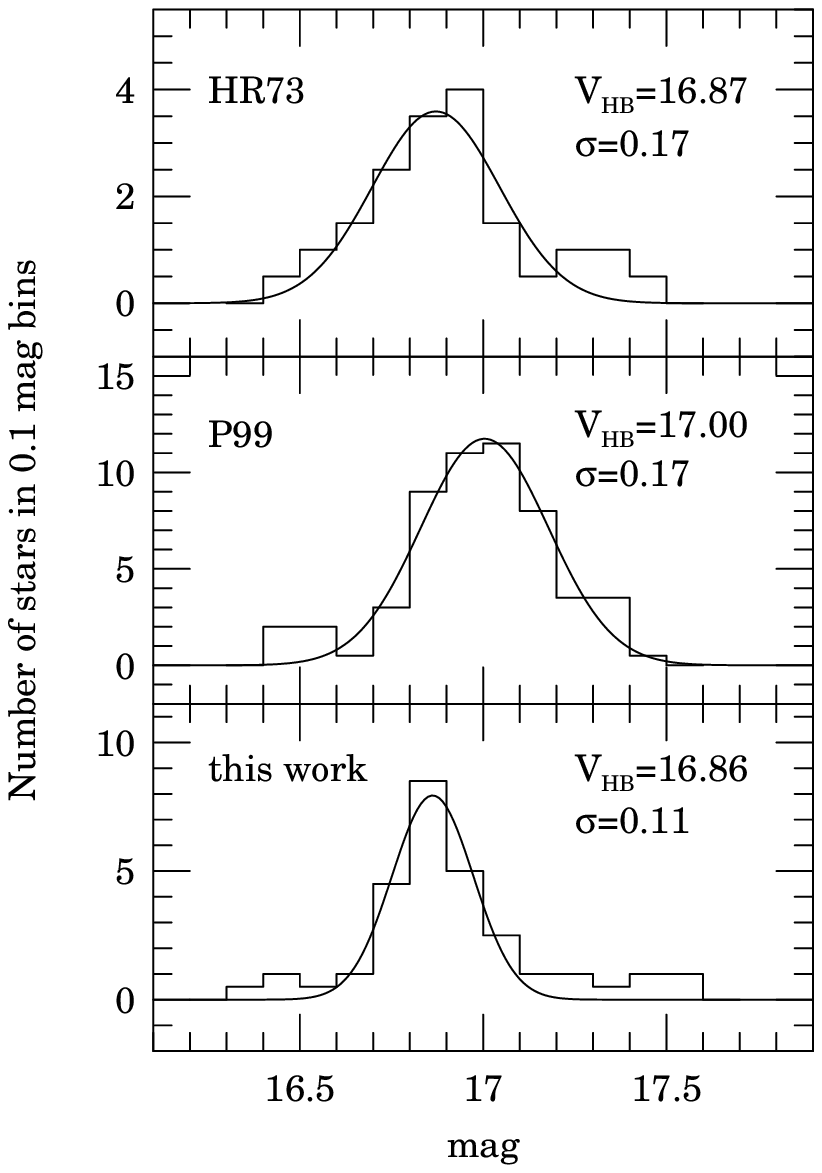}{8.7cm}{0}{100}{100}{-130}{-375}
\caption{The histograms of the distribution in $V$ of
the HB stars in the color interval $0.1<B-V<0.7$  based on color
magnitude diagrams presented by Harris and Racine (1973), Piotto et al
(1999) and in this work.}
\label{fig:hb}
\end{figure}

We conclude that the zero point of the HST-based photometry presented
by Piotto et al. (1999) is shifted by about $0.19\;$mag relatively to
three independent ground-based studies of NGC~6934. That would imply
downward revision of the distance modulus advocated by Piotto et
al.~(1999) to $(m-M)_{V}=16.18\pm 0.05$ if we adopt $V_{HB}=16.86$ as
indicated by our data (the quoted error includes only uncertainty of
measured value of $V_{HB}$).

An alternative way of determining the distance modulus to NGC 6934 is
to compare observed magnitudes of RR~Lyr stars with their absolute
magnitudes. Tables 2 and 3 contain values of $M_{\rm V}$ derived using
3 different methods. By adopting absolute magnitudes derived for RRc
stars with the Kovacs' calibration (column 10th in Table 2) one
obtains $(m-M)_{\rm V}=16.01\pm 0.06$.  In case of Simon \& Clement
calibration (last column in Table 2) we arrive at $(m-M)_{\rm
V}=16.32\pm 0.07$. Finally by using absolute magnitudes derived for
RRab variables with Kovacs \& Jurcsik calibration (Table 3; only stars
with $D_{m}<3$ are considered) we get $(m-M)_{\rm V}=16.07\pm 0.09$.

The most secure calibration of absolute magnitudes of RRab stars
available at the moment is --- in our opinion --- that proposed by
Gould \& Popowski (1998) and based on the statistical-parallax method.
Using purely observational data they derived $M_{\rm V}=0.77\pm 0.13$
at ${\rm [Fe/H]}=-1.60$. After correcting for a 0.01 mag offset
reflecting slightly lower metallicity of the cluster we may adopt
$M_{\rm V}=0.78\pm 0.13$ for RRab stars in NGV~6934 (we used relation
$\Delta M_{\rm V}/\Delta{\rm [Fe/H]}=0.18$ following Fernley et
al. (1997)). From our sample of cluster RRab stars we selected 24
objects with stable light curves and obtained for them an average
value of $<V>=16.873\pm 0.017$.  This in turn leads to an apparent
distance modulus of the cluster $(m-M)_{\rm V}=16.87-0.77=16.10 \pm
0.13$ which is very close to the value $(m-M)_{\rm V}=16.13$ listed in
the catalog of Harris (1996).

Concluding this section we note that four out of five methods used
here to estimate an apparent distance modulus of NGC~6934 give results
which are consistent with each other within quoted errors. Only the
method based on calculation of $M_{\rm V}$ for RRc stars with formulas
given by Simon \& Clement (1993) leads to a relatively larger distance
modulus of the cluster.  By taking an un-weighted mean of four
consistent determinations we obtain $(m-M)_{\rm V}=16.09 \pm 0.06$ for
NGC~6934.

\subsection{W UMa Variables}

We have used the absolute brightness calibration for W UMa-type
binaries (Rucinski 2000) to estimate the absolute magnitudes $M_V$ for
the two newly discovered contact binaries in NGC~6934. Rucinski's
equation gives the relation between $M_V$ at maximum light and the
period, unreddened color $(B-V)_0$ and metallicity:
\begin{equation}
M_V=-4.44\log P + 3.02(B-V)_0 + 0.12
\end{equation}
We adopted $E(B-V)=0.09$ for both stars.  The apparent distance
modulus for each system was calculated as the difference between its
observed $V_{\rm max}$ magnitude and $M_V$ derived from Equation
(17). The last column of Table 4 presents the resulting values of
distance modulus. Both binaries are most likely the foreground
variables because their estimated distance moduli are much smaller
than distance modulus of NGC 6934.

\subsection{Other Variables}

A SX Phe variable NV52 is located in the blue straggler region of the
cluster color-magnitude diagram (see Fig.~6). Large peak-to-peak
amplitude and asymmetrical light curve indicates that NV52 is a
fundamental mode pulsator. Thus we can use the relation of McNamara
(1997) connecting the absolute magnitude of the fundamental SX Phe
pulsators with their periods:
\begin{equation}
M_V=-3.725\cdot\log P_0 - 1.933
\end{equation}
The resulting absolute magnitude of NV52 is $M_V=2.63$ what leads to
estimated distance modulus $(m-M)_V=16.29$. Such a value is consistent
with the assumed cluster membership of the variable.

The variable NV77 is a rather faint and blue object and it possibly
belongs to the group of the cluster blue or yellow stragglers.  It is
hard to say based on our data if there is any periodicity in its light
curve.  The variable is located in the outer part of the cluster but
that location certainly does not preclude cluster membership.
Observed colors suggests that NV77 may be a composite system. We note
that the variable is bluer than the cluster turnoff on the V/B-V
diagram while it is redder than the cluster turnoff on the V/V-I
diagram.

The variable NV85 is located on the red side of the cluster giant
branch and about $0.5\;$mag below its horizontal branch (see Fig.~5).
Our data are best phased with the period $P=1.622\;$days but
$P=2.606\;$days is also possible. Further data are needed to clarify
type of variability and membership status of that star.

The variable NV86 is located at the tip of the cluster red giant
branch.  Our data indicate presence of variations with a period close
to $49\;$days. The star belongs most likely to semiregular, bright
variables, which are common among AGB stars in globular clusters.
 
\section{Conclusions}

We have presented the photometry of 85 variables in the globular
cluster NGC 6934. The photometry was obtained using the newly
developed image subtraction method (Alard \& Lupton 1998; Alard
2000). As first demonstrated by Olech et al.~(1999), image subtraction
method provides a very powerful tool for extracting photometry of
variables located in central regions of globular clusters. As many as
35 variables from our sample are new discoveries. Among these newly
identified stars we have detected 24 RRab stars, 5 RRc stars, 2
eclipsing W UMa systems, one SX Phe star and three other
variables. Our total sample contains photometry for 68 RRab and 10 RRc
stars.

Suntzeff et al.~(1991) estimated that only 6\% of the RR~Lyr variables
hosted by galactic GCs remain to be discovered.  The case of NGC~6934
and recent results obtained by our group for some other clusters (M5,
M55 and NGC~6362) shows that incompleteness of available samples is
much higher. In case of NGC~6934 we have identified 29 new RR~Lyr
stars what amounts up to about 37\% of the total sample for this
cluster.

The periods of RRab variables from NGC 6934 are between 0.4548 and
$0.956\;$days with the mean value of $0.574\;$days. Periods of RRc
stars are between 0.2470 and $0.4008\;$days with the mean at
$0.310\;$days.

Only four RRc stars from our sample have the light curves with quality
good enough for computing their physical parameters from the Fourier
decomposition coefficients. The mean values of the mass, luminosity,
effective temperature and helium abundance for these RRc stars are
$0.63\pm 0.06~M_\odot$, $\log L/L_\odot=1.72\pm 0.01$, $T_{\rm
eff}=7290\pm 27$, and $Y=0.27\pm 0.01$, respectively.

Out of 69 RRab variables only 24 showed stable light curves during our
observations.  For this sub-sample we used the method developed by
Kov\'acs \& Jurcsik (1996, 1997) to derive values of the absolute
magnitude, metallicity and effective temperature which are equal to
$M_V=0.81\pm 0.01$, ${\rm [Fe/H]}=-1.53\pm 0.04$ (Zinn-West scale) and
$T_{\rm eff}=6455\pm 18$, respectively.

From the $B-V$ color at minimum light of the RRab variables we
obtained the color excess to NGC 6934 equal to $E(B-V)=0.09 \pm
0.01$. It is marginally consistent with the recent determination of
Piotto et al. (1999), who derived $E(B-V)=0.05\pm 0.02$, but agrees
well with a value $E(B-V)=0.11$ obtained using the reddening map of
Schlegel et al. (1999).

We obtained five estimates of an apparent distance modulus of the
cluster using various calibrations of $M_{V}$ for RR~Lyr stars and HB
stars. A calibration based on statistical-parallax method (Gould \&
Popowski 1998) which is preferred by us leads to $(m-M)_{\rm
V}=16.10\pm 0.13$.  We noted a likely error in the zero point of
photometry published for the cluster by Piotto et al (1999).

Among cluster RR~Lyr stars we have detected a short period and low
amplitude variable whose characteristic make it a good candidate for
the second overtone pulsator.

Both eclipsing variables detected in this search are W UMa-type
systems and both most likely are the foreground stars.

The magnitude and the color of the single detected SX Phe variable
suggest that it belongs to the cluster blue stragglers group.  Its
distance modulus computed based on relation of McNamara (1997) is
compatible with the cluster membership of that star.

Our sample of newly identified variables includes also one AGB star
with possible period $P\approx 49\;$days, one periodic variable with
$P=1.62\;$days or $P=2.61\;$days and one star which is a likely binary
but for which we have no clue neither for its type of variability nor
for possible periodicity of light variations.

\acknowledgments{We thank Martin Krockenberger and Dimitar Sasselov
for taking significant part of the data used in this paper. We are
indebted to Alex Schwarzenberg-Czerny for his excellent period finding
software and for many stimulating discussions.  We thank Christophe
Alard for his help with ISIS.  JK and AO were supported by the Polish
Committee of Scientific Research through grant 2P03D--003--17 and by
NSF grant AST--9819787 to Bohdan Paczy\'nski.  KZS was
supported by NASA through Hubble Fellowship grant HF-01124.01-A from
the Space Telescope Science Institute, which is operated by the
Association of Universities for Research in Astronomy, Inc., under
NASA contract NAS5-26555.}

\newpage

\begin{footnotesize}
\tablenum{1} 
\begin{planotable}{lcrrrlrrrcl}
\tablewidth{40pc}
\tablecaption{\sc Basic elements of variables from the field of NGC~6934}
\tablehead{\colhead{Star} & \colhead{$\alpha_{J2000.0}$} & \colhead{$\delta_{J2000.0}$} 
& \colhead{$X$} & \colhead{$Y$} & \colhead{$P$} & \colhead{$A_V$} &
\colhead{$<B>$} & \colhead{$<V>$} & \colhead{$<I>$} & \colhead{Type} \\ 
\colhead{} & \colhead{[hh~mm~ss.s]} & \colhead{[$^\circ$~'~"]}
& \colhead{[$\;"\;$]} & \colhead{[$\;"\;$]}  & \colhead{$(days)$} 
& \colhead{} & \colhead{} & \colhead{}  
& \colhead{} & \colhead{} }
\startdata
V1 & 20 34 08.4 & 7 23 39 & $-45$ & $-39$ & 0.56751$^*$ & 0.60 & 17.34 & 16.92 & 16.22 & RRab \nl
V2 & 20 34 08.6 & 7 24 02 & $-40$ & $-14$ & 0.481947 & 1.03 & 17.29 & 16.90 & 16.35 & RRab \nl
V3 & 20 34 11.4 & 7 25 15 & $0$ & $58$ & 0.539806 & 1.09 & 17.22 & 16.85 & 16.28 & RRab \nl
V4 & 20 34 13.9 & 7 25 15 & $39$ & $58$ & 0.616422 & 1.10 & 17.27 & 16.69 & 16.21 & RRab \nl
V5 & 20 34 15.2 & 7 27 58 & $59$ & $221$ & 0.564560 & 0.81 & 17.49 & 17.01 & 16.31 & RRab \nl
V6 & 20 34 09.5 & 7 23 45 & $-27$ & $-33$ & 0.555866 & 1.03 & 17.47 & 16.96 & 16.44 & RRab \nl
V7 & 20 34 17.4 & 7 25 16 & $92$ & $59$ & 0.644049 & 0.68 & 17.29 & 16.82 & 16.19 & RRab \nl
V8 & 20 34 18.0 & 7 25 08 & $100$ & $50$ & 0.623984 & 0.60 & 17.43 & 16.95 & 16.22 & RRab \nl
V9 & 20 34 15.6 & 7 24 36 & $92$ & $59$ & 0.549156 & 0.97 & 17.34 & 16.89 & 16.29 & RRab \nl
V10 & 20 34 02.2 & 7 25 28 & $-135$ & $72$ & 0.519959 & 1.30 & 17.42 & 16.94 & 16.36 & RRab \nl
V11 & 20 34 12.5 & 7 24 46 & $17$ & $28$ & 0.30867$^*$ & 0.50 & 17.13 & 16.88 & 16.40 & RRc \nl
V12 & 20 34 13.2 & 7 23 34 & $29$ & $-44$ & 0.464215 & 1.11 & \nodata & 16.95 & 16.44 & RRab \nl
V13 & 20 34 08.1 & 7 24 42 & $-47$ & $25$ & 0.551334 & 0.87 & 17.36 & 16.94 & 16.32 & RRab \nl
V14 & 20 34 10.9 & 7 22 47 & $-7$ & $-90$ & 0.521990 & 1.00 & 17.42 & 16.85 & 16.32 & RRab \nl
V16 & 20 34 13.7 & 7 24 36 & $36$ & $18$ & 0.604853 & 0.77 & 17.41 & 16.90 & 16.22 & RRab \nl
V17 & 20 34 06.5 & 7 22 30 & $-73$ & $-107$ & 0.598272 & 0.70 & 17.40 & 16.91 & 16.25 & RRab \nl
V18 & 20 34 14.6 & 7 24 09 & $49$ & $-8$ & 0.956070 & 0.47 & 16.98 & 16.50 & 15.79 & RRab \nl
V19 & 20 34 13.2 & 7 24 19 & $30$ & $1$ & 0.480550 & \nodata & \nodata & \nodata & \nodata & RRab \nl
V20 & 20 34 09.6 & 7 24 34 & $-26$ & $17$ & 0.54833$^*$ & 1.02 & 17.23 & 16.78 & 16.23 & RRab \nl
V21 & 20 34 08.9 & 7 24 15 & $-35$ & $-3$ & 0.526829 & 0.69 & 17.45 & 16.94 & 16.40 & RRab \nl
V22 & 20 33 55.3 & 7 21 24 & $-240$ & $-173$ & 0.574280 & 0.50 & 17.47 & 16.93 & 16.30 & RRab \nl
V23 & 20 34 09.3 & 7 24 00 & $-31$ & $-16$ & 0.28643$^*$ & \nodata & \nodata & \nodata & \nodata & RRc \nl
V24 & 20 34 13.8 & 7 23 24 & $37$ & $-53$ & 0.641670 & 0.46 & \nodata & 16.94 & 16.29 & RRab \nl
V25 & 20 34 14.7 & 7 24 54 & $50$ & $37$ & 0.509086 & 1.01 & 17.36 & 16.88 & 16.33 & RRab \nl
V26 & 20 34 13.4 & 7 21 02 & $31$ & $-196$ & 0.259318 & 0.39 & 17.24 & 16.96 & 16.57 & RRc \nl
V27 & 20 34 01.4 & 7 27 39 & $-148$ & $180$ & 0.592204 & 0.84 & 17.46 & 16.88 & 16.21 & RRab \nl
V28 & 20 33 55.6 & 7 25 56 & $-234$ & $100$ & 0.485151 & 1.31 & 17.33 & 16.87 & 16.41 & RRab \nl
V29 & 20 34 05.7 & 7 21 14 & $-85$ & $-183$ & 0.454818 & 1.32 & 17.56 & 17.08 & 16.66 & RRab \nl
V30 & 20 34 22.1 & 7 26 26 & $161$ & $127$ & 0.589853 & 0.79 & 17.43 & 16.92 & 16.25 & RRab \nl
V31 & 20 34 21.1 & 7 22 37 & $146$ & $-101$ & 0.505070 & 1.20 & 17.41 & 16.95 & 16.39 & RRab \nl
V32 & 20 34 10.6 & 7 25 08 & $-10$ & $51$ & 0.511948 & 1.23 & 17.35 & 16.84 & 16.36 & RRab \nl
V33 & 20 34 13.8 & 7 24 29 & $37$ & $12$ & 0.518445 & 1.32 & 17.43 & 16.97 & 16.33 & RRab \nl
V34 & 20 34 09.9 & 7 24 32 & $-21$ & $16$ & 0.560103 & 0.97 & 17.32 & 16.82 & 16.25 & RRab \nl
V35 & 20 34 21.9 & 7 21 56 & $157$ & $-142$ & 0.544222 & 0.76 & 17.48 & 16.99 & 16.32 & RRab \nl
V36 & 20 34 12.1 & 7 23 41 & $10$ & $-35$ & 0.495659 & 1.19 & 17.45 & 16.89 & 16.32 & RRab \nl
V37 & 20 34 12.9 & 7 24 28 & $23$ & $10$ & 0.533186 & 1.20 & \nodata & 16.87 & \nodata & RRab \nl
V38 & 20 34 12.2 & 7 23 59 & $12$ & $-18$ & 0.523562 & 1.04 & 17.18 & 16.71 & \nodata & RRab \nl
V39 & 20 34 11.9 & 7 24 00 & $8$ & $-16$ & 0.502578 & 1.06 & 17.08 & 16.73 & 16.28 & RRab \nl
V40 & 20 34 10.7 & 7 24 44 & $-8$ & $26$ & 0.560755 & 0.96 & \nodata & 16.81 & 16.22 & RRab \nl
V41 & 20 34 13.3 & 7 23 38 & $30$ & $-39$ & 0.520404 & 1.10 & 17.40 & 16.94 & 16.29 & RRab \nl
V42 & 20 34 15.0 & 7 24 39 & $55$ & $20$ & 0.524235 & 1.21 & 17.28 & 16.95 & 16.40 & RRab \nl
V43 & 20 34 12.8 & 7 24 45 & $21$ & $27$ & 0.563218 & 0.92 & 17.32 & 16.83 & 16.25 & RRab \nl
V44 & 20 34 08.5 & 7 23 48 & $-43$ & $-30$ & 0.630384 & 0.50 & 17.39 & 16.90 & 16.22 & RRab \nl
V45 & 20 34 09.2 & 7 24 08 & $-32$ & $-9$ & 0.53660$^*$ & 1.35 & \nodata & 16.97 & 16.35 & RRab \nl
V46 & 20 34 12.3 & 7 23 53 & $14$ & $-24$ & 0.328557 & 0.41 & 17.03 & 16.76 & 16.33 & RRc \nl
V47 & 20 34 12.0 & 7 23 52 & $10$ & $-26$ & 0.640938 & 0.48 & 17.23 & 16.81 & 16.23 & RRab \nl
V48 & 20 34 13.5 & 7 25 08 & $33$ & $52$ & 0.561299 & 0.93 & 17.23 & 16.88 & 16.30 & RRab \nl
V49 & 20 34 12.2 & 7 23 22 & $13$ & $-55$ & 0.285460 & 0.50 & 17.15 & 16.80 & \nodata & RRc \nl
V50 & 20 34 12.4 & 7 23 41 & $15$ & $-37$ & 0.634510 & 0.47 & 17.37 & 16.88 & 16.24 & RRab \nl
V51 & 20 34 11.8 & 7 24 52 & $7$ & $-25$ & 0.564769 & 0.80 & 17.20 & 16.73 & \nodata & RRab \nl
NV52 & 20 34 18.3 & 7 22 14 & $103$ & $-123$ & 0.05976 & 0.46 & 19.27 & 18.92 & 18.48 &SX Phe \nl
NV53 & 20 34 13.6 & 7 24 00 & $33$ & $-16$ & 0.28235 & 0.49 & 17.16 & 16.88 & 16.45 & RRc \nl
NV54 & 20 34 12.6 & 7 24 35 & $30$ & $10$ & 0.59020 & 0.73 & 17.27 & 16.79 & 16.19 & RRab \nl
NV55 & 20 34 13.3 & 7 24 27 & $18$ & $123$ & 0.77828 & 0.25 & 17.15 & 16.66 & 15.98 & RRab \nl
NV56 & 20 34 12.5 & 7 24 18 & $18$ & $1$ & 0.29104 & 0.46 & 17.02 & 16.64 & \nodata & RRc \nl
NV57 & 20 34 12.4 & 7 24 10 & $15$ & $-7$ & 0.68712 & \nodata & \nodata & \nodata & \nodata & RRab \nl
NV58 & 20 34 12.1 & 7 25 05 & $11$ & $47$ & 0.40082 & 0.47 & 17.04 & 16.69 & 16.15 & RRc \nl
NV59 & 20 34 12.0 & 7 24 15 & $10$ & $-2$ & 0.53855 & \nodata & \nodata & \nodata & \nodata & RRab \nl
NV60 & 20 34 12.0 & 7 24 56 & $10$ & $38$ & 0.66040 & 0.40 & 17.31 & 16.85 & 16.20 & RRab \nl
NV61 & 20 34 11.9 & 7 23 10 & $8$ & $-66$ & 0.528? & 0.38 & 17.24 & 16.87 & 16.29 & RRab? \nl
NV62 & 20 34 11.7 & 7 24 26 & $6$ & $9$ & 0.53067 & \nodata & \nodata & \nodata & \nodata & RRab \nl
NV63 & 20 34 11.6 & 7 24 26 & $3$ & $9$ & 0.57564 & \nodata & \nodata & \nodata & \nodata & RRab \nl
NV64 & 20 34 11.4 & 7 24 18 & $1$ & $1$ & 0.57102 & \nodata & \nodata & \nodata & \nodata & RRab \nl
NV65 & 20 34 11.4 & 7 24 15 & $1$ & $-2$ & 0.65905 & \nodata & \nodata & \nodata & \nodata & RRab \nl
NV66 & 20 34 11.1 & 7 24 16 & $-3$ & $-1$ & 0.54078 & \nodata & \nodata & \nodata & \nodata & RRab \nl
NV67 & 20 34 10.9 & 7 23 55 & $-6$ & $-21$ & 0.61333 & \nodata & \nodata & \nodata & \nodata & RRab \nl
NV68 & 20 34 10.9 & 7 24 00 & $-7$ & $-17$ & 0.33534 & \nodata & \nodata & \nodata & \nodata & RRc \nl
NV69 & 20 34 10.8 & 7 23 15 & $-8$ & $-61$ & 0.24700 & 0.09 & 17.17 & 16.90 & 16.57 & RRe? \nl
NV70 & 20 34 10.7 & 7 24 06 & $-9$ & $-10$ & 0.53935 & \nodata & \nodata & \nodata & \nodata & RRab \nl
NV71 & 20 34 10.7 & 7 23 54 & $-9$ & $-23$ & 0.57269 & \nodata & \nodata & \nodata & \nodata & RRab \nl
NV72 & 20 34 10.5 & 7 25 20 & $-12$ & $61$ & 0.66785 & 0.29 & 17.29 & 16.85 & 16.18 & RRab \nl
NV73 & 20 34 09.8 & 7 24 47 & $-22$ & $29$ & 0.50621 & 0.51 & 17.29 & 16.92 & 16.42 & RRab \nl
NV74 & 20 34 09.3 & 7 24 08 & $-30$ & $-9$ & 0.56813 & 1.00 & 17.35 & 16.86 & 16.23 & RRab \nl
NV75 & 20 34 02.8 & 7 19 35 & $-128$ & $-288$ & 0.28207 & 0.36 & 17.78 & 17.15 & 16.23 & EW \nl
NV76 & 20 33 54.6 & 7 19 50 & $-251$ & $-272$ & 0.33649 & 0.31 & 18.23 & 17.94 & 17.58 & EW \nl
NV77 & 20 34 10.5 & 7 24 24 & $236$ & $47$ & \nodata & ? & 19.43 & 19.10 & 18.36 & LP \nl
NV78 & 20 34 12.1 & 7 24 38 & $12$ & $21$ & 0.54230 & \nodata & \nodata & \nodata & \nodata & RRab \nl
NV79 & 20 34 10.5 & 7 24 24 & $-12$ & $7$ & 0.62187 & \nodata & \nodata & \nodata & \nodata & RRab \nl
NV80 & 20 34 11.7 & 7 24 07 & $6$ & $-10$ & 0.54427 & \nodata & 17.25 & 16.77 & \nodata & RRab \nl
NV81 & 20 34 11.2 & 7 24 23 & $-2$ & $6$ & 0.57262 & \nodata & \nodata & \nodata & \nodata & RRab \nl
NV82 & 20 34 10.7 & 7 24 17 & $-9$ & $0$ & 0.73113 & \nodata & \nodata & \nodata & \nodata & RRab \nl
NV83 & 20 34 11.2 & 7 24 26 & $-1$ & $8$ & 0.54055 & \nodata & \nodata & \nodata & \nodata & RRab \nl
NV84 & 20 34 12.1 & 7 24 28 & $12$ & $11$ & 0.66535 & \nodata & \nodata & \nodata & \nodata & RRab \nl
NV85 & 20 34 31.0 & 7 21 57 & $294$ & $-137$ & 1.622   & 0.13 & 18.34 & 17.46 & 16.33 & ?  \nl
NV86 & 20 34 19.5 & 7 22 51 & $122$ & $-85$ & $\sim$49 & 0.15 & 15.53 & 13.90 & 12.03 & LP
\enddata 
\label{tab:var}
\end{planotable}
\end{footnotesize}

\begin{footnotesize}
\tablenum{2} 
\begin{planotable}{lrrrrrrrrrr}
\tablewidth{40pc}
\tablecaption{\sc Parameters for the RRc variables in NGC 6934.}
\tablehead{\colhead{Star} & \colhead{$A_0$} & \colhead{$A_1$} 
& \colhead{$\phi_{21}$} & \colhead{$\phi_{31}$} & \colhead{$M$} & \colhead{$\log{L}$} &
\colhead{$T_{\rm eff}$} & \colhead{$Y$} & \colhead{$M_V^{Ko}$} & \colhead{$M_V$} \\
\colhead{} & \colhead{} & \colhead{} 
& \colhead{} & \colhead{} & \colhead{[$\;M_{\odot}\;$]} & \colhead{} &
\colhead{[$\;K\;$]} & \colhead{} & \colhead{} & \colhead{} }
\startdata
V11 & 16.895 & 0.259 & 3.076 & 6.077 & 0.633 & 1.709 & 7335 & 0.272 & 0.747 & 0.460\nl
& $\pm$0.000 & $\pm$0.002 & $\pm$0.040 & $\pm$0.070 & $\pm$0.011 &$\pm$0.004 &  $\pm$6 &$\pm$0.002 &$\pm$0.007& \nl
V26 & 16.634 & 0.218 & 3.027 & 6.039 & 0.620 & 1.684 & 7391 & 0.279 & 0.765 & 0.8727\nl
& $\pm$0.000 & $\pm$0.002 & $\pm$0.058 & $\pm$0.211 & $\pm$0.027 &$\pm$0.012 &  $\pm$20 &$\pm$0.007 &$\pm$0.008& \nl
V46 & 16.761 & 0.209 & 2.734 & 6.234 & 0.629 & 1.728 & 7280 & 0.267 & 0.791 & 0.412\nl
& $\pm$0.000 & $\pm$0.002 & $\pm$0.111 & $\pm$0.120 & $\pm$0.019 &$\pm$0.007 &  $\pm$11 &$\pm$0.003 &$\pm$0.011& \nl
V49 & 16.817 & 0.234 & 2.804 & 5.135 & 0.772 & 1.728 & 7325 & 0.261 & 0.767 & 0.412\nl
& $\pm$0.000 & $\pm$0.003 & $\pm$0.080 & $\pm$0.101 & $\pm$0.020 &$\pm$0.006 &  $\pm$9 &$\pm$0.003 &$\pm$0.013& \nl
NV53 & 16.878 & 0.241 & 2.896 & 5.861 & 0.639 & 1.681 & 7406 & 0.279 & 0.816 & 0.530\nl
& $\pm$0.000 & $\pm$0.004 & $\pm$0.082 & $\pm$0.278 & $\pm$0.045 &$\pm$0.016 &  $\pm$26 &$\pm$0.008 &$\pm$0.018& \nl
NV56 & 16.634 & 0.218 & 3.027 & 6.039 & 0.620 & 1.684 & 7391 & 0.279 & 0.765 & 0.522\nl
& $\pm$0.000 & $\pm$0.009 & $\pm$0.529 & $\pm$1.024 & $\pm$0.161 &$\pm$0.059 &  $\pm$97 &$\pm$0.029 &$\pm$0.081& \nl
NV58 & 16.698 & 0.231 & 4.193 & 7.807 & 0.468 & 1.726 & 7218 & 0.276 & 0.649 & 0.417\nl
& $\pm$0.000 & $\pm$0.001 & $\pm$0.112 & $\pm$0.137 & $\pm$0.016 &$\pm$0.008 &  $\pm$12 &$\pm$0.004 &$\pm$0.009& \nl
NV69 & 16.897 & 0.045 & 3.541 & 3.678 & 1.036 & 1.747 & 7341 & 0.247 & 0.859 & 0.365\nl
& $\pm$0.000 & $\pm$0.001 & $\pm$0.483 & $\pm$1.268 & $\pm$0.333 &$\pm$0.074 &  $\pm$119 &$\pm$0.032 &$\pm$0.022
\enddata 
\label{tab:rrc}
\end{planotable}
\end{footnotesize}

\begin{footnotesize}
\tablenum{3} 
\begin{planotable}{lrrrrrrrrrrr}
\tablewidth{40pc}
\tablecaption{\sc Parameters for the RRab Variables in NGC 6934}
\tablehead{\colhead{Star} & \colhead{$P$} & \colhead{$A_0$} & \colhead{$A_1$} 
& \colhead{$\phi_{31}$} & \colhead{$\phi_{41}$} & \colhead{$M_V$} & \colhead{$\sigma_{M_V}$} &
\colhead{[Fe/H]} & \colhead{$\sigma_{\rm [Fe/H]}$} & \colhead{$T_{\rm eff}$} & \colhead{$D_m$} \\
\colhead{} & \colhead{[\/days\/]} & \colhead{} 
& \colhead{} & \colhead{} & \colhead{} & \colhead{} &
\colhead{} & \colhead{} & \colhead{} & \colhead{[$\;K\;$]} & \colhead{} }
\startdata
V1  & 0.567511 & 16.903 & 0.255 & 4.810 & 1.259 & 0.807 & 0.176 & $-$1.630 & 0.992 & 6353 & 3.87\nl
V2  & 0.481947 & 16.933 & 0.381 & \nodata & \nodata & \nodata & \nodata & \nodata & \nodata & \nodata & 14.49\nl
V3  & 0.539850 & 16.894 & 0.376 & 4.735 & 1.195 & 0.780 & 0.081 & $-$1.581 & 0.099 & 6439 & 2.49\nl
V4  & 0.616310 & 16.690 & 0.345 & 5.069 & 1.277 & 0.723 & 0.089 & $-$1.544 & 0.174 & 6386 & 2.96\nl
V5  & 0.564507 & 16.921 & 0.279 & 5.158 & 1.590 & 0.836 & 0.091 & $-$1.146 & 0.235 & 6447 & 2.15\nl
V6  & 0.555866 & 16.951 & 0.344 & 4.884 & 1.342 & 0.789 & 0.084 & $-$1.468 & 0.064 & 6429 & 1.77\nl
V7  & 0.644068 & 16.857 & 0.255 & 5.434 & 2.059 & 0.765 & 0.103 & $-$1.203 & 0.510 & 6333 & 1.11\nl
V8  & 0.623984 & 16.869 & 0.203 & \nodata & \nodata & \nodata & \nodata & \nodata & \nodata & \nodata & 21.67\nl
V9  & 0.549104 & 16.912 & 0.330 & 4.729 & 1.080 & 0.789 & 0.082 & $-$1.639 & 0.068 & 6412 & 2.59\nl
V10 & 0.519949 & 16.895 & 0.439 & 4.717 & 1.118 & 0.776 & 0.079 & $-$1.498 & 0.100 & 6506 & 1.95\nl
V12 & 0.464215 & 16.968 & 0.403 & 4.661 & 0.958 & 0.865 & 0.078 & $-$1.272 & 0.057 & 6591 & 1.16\nl
V13 & 0.551334 & 16.956 & 0.300 & 5.263 & 1.913 & 0.855 & 0.092 & $-$0.933 & 0.185 & 6481 & 3.97\nl
V14 & 0.522084 & 16.898 & 0.398 & \nodata & \nodata & \nodata & \nodata & \nodata & \nodata & \nodata & 61.88\nl
V16 & 0.604817 & 16.863 & 0.261 & 5.291 & 1.837 & 0.803 & 0.093 & $-$1.184 & 0.052 & 6383 & 1.73\nl
V17 & 0.598272 & 16.902 & 0.260 & 5.162 & 1.880 & 0.799 & 0.091 & $-$1.322 & 0.092 & 6349 & 1.34\nl
V18 & 0.955969 & 16.512 & 0.196 & \nodata & \nodata & \nodata & \nodata & \nodata & \nodata & \nodata & 38.63\nl
V20 & 0.548327 & 16.786 & 0.346 & 4.829 & 1.187 & 0.793 & 0.083 & $-$1.501 & 0.090 & 6441 & 1.70\nl
V21 & 0.526829 & 16.925 & 0.243 & \nodata & \nodata & \nodata & \nodata & \nodata & \nodata & \nodata & 16.14\nl
V22 & 0.574282 & 16.927 & 0.242 & \nodata & \nodata & \nodata & \nodata & \nodata & \nodata & \nodata & 29.84\nl
V24 & 0.641670 & 16.942 & 0.176 & \nodata & \nodata & \nodata & \nodata & \nodata & \nodata & \nodata & 18.01\nl
V25 & 0.509048 & 16.861 & 0.376 & 4.673 & 1.008 & 0.817 & 0.079 & $-$1.499 & 0.052 & 6494 & 3.36\nl
V27 & 0.592221 & 16.870 & 0.297 & 5.299 & 1.749 & 0.804 & 0.092 & $-$1.105 & 0.058 & 6433 & 1.89\nl
V28 & 0.485111 & 16.886 & 0.483 & 4.734 & 1.168 & 0.806 & 0.078 & $-$1.287 & 0.068 & 6591 & 2.24\nl
V29 & 0.454802 & 17.014 & 0.470 & 4.756 & 1.101 & 0.856 & 0.078 & $-$1.094 & 0.133 & 6656 & 1.62\nl
V30 & 0.589850 & 16.876 & 0.275 & 5.172 & 1.795 & 0.804 & 0.090 & $-$1.264 & 0.045 & 6383 & 0.84\nl
V31 & 0.505063 & 16.966 & 0.400 & 4.788 & 1.124 & 0.823 & 0.081 & $-$1.322 & 0.069 & 6538 & 0.76\nl
V32 & 0.511934 & 16.882 & 0.420 & 4.818 & 1.174 & 0.807 & 0.081 & $-$1.320 & 0.055 & 6539 & 1.08\nl
V33 & 0.518728 & 16.822 & 0.407 & \nodata & \nodata & \nodata & \nodata & \nodata & \nodata & \nodata & 14.74\nl
V34 & 0.560108 & 16.839 & 0.323 & \nodata & \nodata & \nodata & \nodata & \nodata & \nodata & \nodata & 5.11\nl
V35 & 0.544222 & 16.955 & 0.290 & 5.208 & 1.603 & 0.864 & 0.091 & $-$0.969 & 0.193 & 6504 & 4.65\nl
V36 & 0.495659 & 16.829 & 0.384 & 4.771 & 1.015 & 0.842 & 0.081 & $-$1.294 & 0.095 & 6554 & 4.22\nl
V37 & 0.533173 & 16.858 & 0.345 & 5.053 & 1.578 & 0.837 & 0.087 & $-$1.118 & 0.144 & 6501 & 1.94\nl
V38 & 0.523553 & 16.775 & 0.396 & \nodata & \nodata & \nodata & \nodata & \nodata & \nodata & \nodata & 12.68\nl
V39 & 0.502574 & 16.750 & 0.357 & 5.116 & 1.572 & 0.881 & 0.089 & $-$0.868 & 0.241 & 6584 & 2.53\nl
V40 & 0.560755 & 16.843 & 0.339 & 4.920 & 1.257 & 0.788 & 0.085 & $-$1.445 & 0.061 & 6438 & 1.91\nl
V41 & 0.520404 & 16.987 & 0.378 & \nodata & \nodata & \nodata & \nodata & \nodata & \nodata & \nodata & 10.42\nl
V42 & 0.524235 & 16.984 & 0.406 & 4.814 & 0.973 & 0.796 & 0.082 & $-$1.390 & 0.118 & 6530 & 4.36\nl
V43 & 0.563180 & 16.846 & 0.317 & 4.938 & 1.344 & 0.797 & 0.086 & $-$1.435 & 0.056 & 6420 & 1.63\nl
V44 & 0.630373 & 16.902 & 0.193 & \nodata & \nodata & \nodata & \nodata & \nodata & \nodata & \nodata & 13.87 \nl
V45 & 0.536598 & 16.914 & 0.532 & \nodata & \nodata & \nodata & \nodata & \nodata & \nodata & \nodata & 21.12\nl
V47 & 0.640886 & 16.801 & 0.181 & 5.563 & 2.159 & 0.819 & 0.099 & $-$1.013 & 0.120 & 6337 & 2.78\nl
V48 & 0.561299 & 16.901 & 0.319 & 4.988 & 1.430 & 0.804 & 0.086 & $-$1.356 & 0.044 & 6432 & 0.63\nl
V50 & 0.634531 & 16.890 & 0.174 & \nodata & \nodata & \nodata & \nodata & \nodata & \nodata & \nodata & 38.68\nl
V51 & 0.564785 & 16.744 & 0.257 & \nodata & \nodata & \nodata & \nodata & \nodata & \nodata & \nodata & 8.72\nl
NV54 & 0.590199 & 16.799 & 0.253 & 5.072 & 1.635 & 0.804 & 0.090 & $-$1.400 & 0.121 & 6356 & 1.89\nl
NV55 & 0.778281 & 16.656 & 0.112 & \nodata & \nodata & \nodata & \nodata & \nodata & \nodata & \nodata & 20.39\nl
NV60 & 0.660405 & 16.857 & 0.165 & \nodata & \nodata & \nodata & \nodata & \nodata &  \nodata & \nodata & 29.07\nl
NV72 & 0.667850 & 16.851 & 0.133 & 5.751 & 2.200 & 0.823 & 0.103 & $-$0.906 & 0.112 & 6323 & 4.68\nl
NV73 & 0.506209 & 16.923 & 0.152 & \nodata & \nodata & \nodata & \nodata & \nodata & \nodata & \nodata & 11.13\nl
NV74 & 0.568126 & 16.782 & 0.412 & \nodata & \nodata & \nodata & \nodata & \nodata & \nodata & \nodata & 14.23\nl
NV80 & 0.544275 & 16.768 & 0.302 & \nodata & \nodata & \nodata & \nodata & \nodata & \nodata & \nodata & 8.90
\enddata 
\label{tab:rrab} 
\end{planotable}
\end{footnotesize}

\tablenum{4} 
\begin{planotable}{lccccc}
\tablewidth{27pc} 
\tablecaption{\sc Basic elements of W UMa variables in NGC~6934}
\tablehead{\colhead{Star} & \colhead{$P$} & \colhead{$V_{\rm max}$} & 
\colhead{$B_{\rm max}$} & \colhead{$(B-V)_{\rm max}$} & \colhead{$(m-M)_V$} \\ 
\colhead{} & \colhead{[\/days\/]} & \colhead{} & \colhead{} & \colhead{} &
\colhead{} } 
\startdata 
NV75 & 0.28207 & 16.99 & 17.62 & 0.63 & 12.82 \nl 
NV76 & 0.33649 & 17.80 & 18.08 & 0.28 & 15.02 
\enddata
\label{tab:wuma} 
\end{planotable}

\end{document}